%% file: main.tex
\documentclass[twocolumn]{aastex62} 

\input preamble.tex
\newcommand*{\myfont}{\fontfamily{phv}\selectfont}

\shorttitle{GDIGS Overview}
\shortauthors{Anderson et al.}

\begin{document}

\title{The GBT Diffuse Ionized Gas Survey (GDIGS): Survey Overview and First Data Release}

\author[0000-0001-8800-1793]{L.~D.~Anderson}
\affiliation{Department of Physics and Astronomy, West Virginia University, Morgantown, WV 26506, USA}
\affiliation{Adjunct Astronomer at the Green Bank Observatory, P.O. Box 2, Green Bank, WV 24944, USA}
\affiliation{Center for Gravitational Waves and Cosmology, West Virginia University, Chestnut Ridge Research Building, Morgantown, WV 26505, USA}

\author[0000-0001-8061-216X]{Matteo~Luisi}
\affiliation{Department of Physics and Astronomy, West Virginia University, Morgantown, WV 26506, USA}
\affiliation{Center for Gravitational Waves and Cosmology, West Virginia University, Chestnut Ridge Research Building, Morgantown, WV 26505, USA}

\author[0000-0002-1311-8839]{Bin Liu}
\affiliation{National Astronomical Observatories, Chinese Academy of Sciences, Beijing 100012, China}
\affiliation{CAS Key Laboratory of FAST, NAOC, Chinese Academy of Sciences}
\affiliation{Department of Physics and Astronomy, West Virginia University, Morgantown, WV 26506, USA}

\author[0000-0003-0640-7787]{Trey~V.~Wenger}
\affiliation{Dominion Radio Astrophysical Observatory, Herzberg Astronomy and Astrophysics Research Centre, National Research Council, P.O. Box 248, Penticton, BC V2A 6J9, Canada}

\author[0000-0002-2465-7803]{Dana.~S.~Balser}
\affiliation{National Radio Astronomy Observatory, 520 Edgemont Road, Charlottesville, VA 22903, USA}

\author[0000-0003-4866-460X]{T.~M.~Bania}
\affiliation{Institute for Astrophysical Research, Astronomy Department, Boston University, 725 Commonwealth Ave., Boston, MA 02215, USA}

\author[0000-0002-9947-6396]{L.~M.~Haffner}
\affiliation{Department of Physical Sciences, Embry-Riddle Aeronautical University, Daytona Beach FL 32114, USA}

\author[0000-0002-4727-7619]{Dylan~J.~Linville}
\affiliation{Department of Physics and Astronomy, West Virginia University, Morgantown, WV 26506, USA}
\affiliation{Center for Gravitational Waves and Cosmology, West Virginia University, Chestnut Ridge Research Building, Morgantown, WV 26505, USA}

\author[0000-0002-3758-2492]{J.~L.~Mascoop}
\affiliation{Department of Physics and Astronomy, West Virginia University, Morgantown, WV 26506, USA}
\affiliation{Center for Gravitational Waves and Cosmology, West Virginia University, Chestnut Ridge Research Building, Morgantown, WV 26505, USA}

\correspondingauthor{L.D.~Anderson}
\email{loren.anderson@mail.wvu.edu}

\begin{abstract}
The Green Bank Telescope (GBT) Diffuse Ionized Gas Survey (GDIGS) traces 
ionized gas in the Galactic midplane by measuring $4-8$\,\ghz\ radio recombination line (RRL) emission. The nominal survey zone is $32.3\degree>\ell >-5\degree$, $\absb<0.5\degree$, but coverage extends above and below the plane in select fields, and additionally includes the areas around W47 ($\ell \simeq 37.5\degree$) and W49 ($\ell \simeq 43\degree$).  
GDIGS simultaneously observes 22 \hna\ (15 usable), 25 \hnb\ (18 usable), and 8 \hng\ RRLs (all usable), as well as
multiple molecular line transitions (including of \formr, \form, and \methanol).
Here, we describe the GDIGS survey parameters and characterize the RRL data, focusing primarily on the \hna\ data.
We produce
sensitive data cubes by averaging the usable RRLs, after
first smoothing to a common spectral resolution of 0.5\,\kms\ and a
spatial resolution of 2\arcmper65 for \hna, 2\arcmper62 for \hnb, and 2\arcmper09 for
\hng.
The average spectral noise per spaxel in the \hna\ data cubes is $\sim\!10$\,mK ($\sim\!5$\,m\!\jyb).
This sensitivity allows GDIGS to detect RRLs from plasma throughout the inner Galaxy.  The GDIGS \hna\ data are sensitive to emission measures EM\,$\gtrsim 1100$\,cm$^{-6}\,\pc$, which corresponds to a mean electron density $\langle n_e \rangle \gtrsim 30\,\percc$ for a 1\,pc path length or
$\langle n_e \rangle \gtrsim 1\,\percc$ for a 1\,kpc path length.
\end{abstract}


\section{Introduction}
The myriad Galactic plane surveys undertaken over the last decades
have given us a clear view of nearly all the components of the Milky
Way disk.  These surveys cover the visible through radio regimes,
tracing stars (2MASS, UKIDSS), warm dust (\textit{MSX}, {\it Spitzer} GLIMPSE
and MIPSGAL), cold dust ({\it Herschel} Hi-Gal, AKARI, ATLASGAL,
BGPS), compact ionized gas (SUMSS, MAGPIS, CORNISH), \hi\ (IGPS), and molecular gas
(e.g., GRS, HOPS, SPLASH).  The major omission, however, is data that
are sensitive to $10^4\,\K$ ``diffuse ionized gas'' (DIG) in the Galactic plane. 
The DIG, sometimes called the ``warm ionized medium'' or ``WIM'' is a low-density plasma whose existence was first proposed by \citet{hoyle63}.  It is a major
component of the interstellar medium (ISM), making up $\sim 20\%$ of the total Milky Way gas
mass and $\gtrsim 90\%$ of its ionized gas mass \citep{reynolds91a}, and providing a major source of pressure at the midplane
\citep{boulares90}.  
Although
we have known about the DIG for over half a century and it is a
major component of the ISM, there remain major
unanswered questions regarding its origin, distribution, and
characteristics.
We
cannot therefore fully trace the recycling of material that takes place in
the ISM, nor can we develop a complete picture of how star formation impacts the ISM of our Galaxy.

The DIG exists in a variety of environments and hence has a range of densities.  
In the immediate vicinity of an \hii\ region, but outside its photodissociation region (PDR), there often
exists an ionized gas ``halo'' \citep{anantharamaiah85, anantharamaiah86}.  These halos may be due to
photons leaking through the \hii\ region PDRs, as has been shown for the \hii\ regions RCW120  
\citep{anderson15a} and NGC7538 
\citep{luisi16}.  \citet{pellegrini20} modeled the emission from \hii\ regions in a Milky Way-like galaxy and found that their halos were bright due to leakage.  There also is widespread lower density ionized gas in the Milky Way disk that cannot definitively be associated with any individual \hii\ region \citep[e.g.,][]{geyer18}.  Here, we refer to all diffuse ionized gas outside of \hii\ regions as the ``DIG.''

Extragalactic studies of the DIG have benefited from recent hardware advances that allow for two-dimensional mapping using integral field units such as SAURON \citep{bacon01}, SparsePak \citep{bershady05}, PMAS
\citep{roth05}, VIRUS-P \citep{hill08}, MUSE \citep{mcdermid08}, and MaNGA \citep{bundy15}.  These instruments can map the emission from galaxies quickly and determine their ionized gas properties.  The forthcoming Local Volume Mapper \citep{kollmeier17} of the Sloan Digital Sky Survey \citep[SDSS;][]{gunn06} will allow for $\sim2\times 10^7$ simultaneous spectra to be taken, allowing plasma diagnostics over a range of physical scales.

Most studies of the Milky Way DIG have been conducted by observing H$\alpha$ emission.
Although H$\alpha$ is very bright compared to other ionized gas
tracers, it suffers from extinction.
This drastically limits the distance to which the DIG in
the inner Galaxy can be studied.
The all-sky H$\alpha$
WHAM survey \citep{reynolds98, haffner03} gives us a clear view of the global
properties of diffuse ionized gas.  Due to extinction and its relatively coarse
resolution of $\sim1\degree$, however, WHAM data cannot be used for
detailed studies of the inner Galaxy mid-plane where significant 
diffuse ionized gas resides.

The Milky Way DIG can also be studied using
far-infrared (FIR) fine-structure lines.  Unlike H$\alpha$, such lines do not suffer from extinction.  To date, however, such studies have been limited to select sight lines or coarse maps, often at poor spectral resolution.  For example the only large-scale map extant is that of the \nii\ 122 and 205\,\micron\ lines, mapped by the FIRAS instrument on the Cosmic Background Explorer ({\it COBE}\,) at 7\degree\ angular resolution and $\sim\!700\,\kms$ spectral resolution \citep[][]{wright91}.  
Since the ionization potential of nitrogen is greater than that of hydrogen, the distribution of ionized nitrogen should be similar to that of ionized hydrogen; the lines of nitrogen are therefore the most useful FIR DIG tracers.
\citet{goldsmith15} used {\it Herschel} observations of the 122 and 205\,\micron\  \nii\ fine-structure lines along $\sim\!150$
lines of sight in the plane.  They found electron densities
$n_e\simeq 10-50$\,\percc, and coined the phase for this plasma the ``dense warm ionized gas,'' D-WIM. \citet{pineda19} extended this work 
to the analysis of 11 lines of sight in the plane, and found electron densities $n_e \simeq 10-200$\,\percc.
The electron densities derived in these studies are higher than those of extraplanar DIG \citep[which has $n_e \simeq 0.1\,\percc$;][]{haffner09} and lower than those of most \hii\ regions \citep[e.g., the Orion nebula has $n_e\simeq 10^{2.3}-10^{4.5}$\,\percc;][]{lockman75}.

Observations of radio recombination lines (RRLs) give us an
opportunity to investigate the Galactic mid-plane DIG distribution
throughout the Galactic disk at high spatial and spectral resolution.  RRLs are produced from recombining ions
and electrons and are the higher quantum number corollaries to
H$\alpha$ emission.  We therefore expect RRL emission from \hii\ regions surrounding OB stars and from diffuse ionized gas.  There are, however, few large-scale RRL surveys.  Past RRL observations of the Galactic DIG \citep[e.g.,][]{roshi00,
  roshi01, baddi12} found prevalent emission, but are inappropriate for detailed studies of
the DIG due to poor angular resolution.
The beam-sampled RRL survey SIGGMA has a sensitivity of $\sim\!1$\,m\!\jyb at 5\,\kms\ spectral resolution and $\sim\!3\arcmper4$ spatial
resolution \citep{liu13, liu19}.
SIGGMA is sensitive to discrete sources but due to the observing strategy has decreased sensitivity to extended emission over $\sim\!1\degree$ in extent.  
In the only other large-scale RRL survey extant, \citet{alves15} report 6.4\,m\!\jyb\ sensitivity at
20\,\kms\ spectral resolution.  Both of these comparison surveys are near $1.4\,\ghz$.

In previous RRL observations by our group, we serendipitously discovered
prevalent, diffuse ionized gas in the Milky Way.  We created the
\hii\ Region Discovery Survey \citep[HRDS;][]{bania10}, a RRL survey
of discrete \hii\ regions throughout the Galactic disk.  To date, we
have discovered $\sim~\!\!\!1400$ nebulae, approximately doubling the
previously-known sample in the Galaxy.  Nearly 30\% of all observed
positions, however, have two or more distinct RRL velocities.  We
showed using subsequent observations that one of these components was
from the DIG and the other was from the compact \hii\ regions that we
targeted \citep{anderson15b}.  The multiple-velocity \hii\ regions are
rare at $\ell > 32\degree$, implying less diffuse ionized gas there.  Our observations showed that diffuse
ionized gas in the Galactic plane is easily detected by the Green Bank
Telescope (GBT).  With its 100-m diameter and unblocked aperture, the
GBT has extraordinary surface brightness sensitivity, making it the
ideal instrument for further studies of the DIG.

Here, we report on a new fully-sampled survey of RRL emission, the GBT Diffuse Ionized Gas Survey (GDIGS).  
First GDIGS data on the giant \hii\ region W43 were published by \citet{luisi20}.
The same observing mode was used to study the ionized gas of S235 by \citet{anderson19b}.  This paper describes the data acquisition and reduction methodology and characterizes the GDIGS RRL data.


\section{GDIGS Data Acquisition and Processing}
GDIGS is optimized to provide sensitive  RRL data of the DIG, at good spatial resolution. We use the GBT C-band receiver, which operates from 4--8\,\ghz, in total power mode.  The C-band receiver can tune to the largest number of usable RRLs of any GBT receiver, at spatial resolutions of $\sim\!2\arcmin$.
The GDIGS coverage area includes the bulk of the midplane DIG emission observable by the GBT.  
We characterize the GDIGS RRL survey parameters in Table~\ref{tab:survey}, which summarizes much of the discussion below.

\begin{deluxetable}{ll}
\tablecaption{GDIGS RRL Survey Parameters\label{tab:survey}}
\tablehead{\colhead{\!\!\!\!\!\!\!\!\!\!\!\!\!\!\!\!\!\!\!\!\!\!\!\!\!Observing Parameters}}
\startdata
Observing Dates & 7/2016 -- 12/2019\\
Nominal Coverage Area & $32.3\degree>\ell >-5\degree$, $\absb<0.5\degree$\\
Usable frequencies & $4.7$--$7.3\,\ghz$\\\\
Mean \hna\ frequency & 5.7578\,\ghz \\
\hna\ spatial resolution & $2\arcmper65$\\
Integration time per \hna\ beam & 45\,s \\\\
Mean \hnb\ frequency & 5.7959\,\ghz \\
\hnb\ spatial resolution & $2\arcmper62$\\
Integration time per \hnb\ beam & 44\,s \\\\
Mean \hng\ frequency & 6.4528\,\ghz \\
\hng\ spatial resolution & $2\arcmper09$\\
Integration time per \hng\ beam & 28\,s \\\\
Data products & \\\hline
LSR Velocity Range & $-300$ to $+300\,\kms$\\
Pixel size & $30\arcsec$\\
Channel spacing & 0.5\,\kms\\
Median spectral \hna\ rms noise & 10.3\,mK\\
Median spectral \hnb\ rms noise & 10.8\,mK\\
Median spectral \hng\ rms noise & 21.3\,mK
\enddata
\end{deluxetable}

The GDIGS data are taken in intensity units of antenna temperature ($T_A$, in \K).  
At C-band, atmospheric effects are negligible and therefore no correction for atmospheric opacity is required.  The
unblocked aperture of the GBT provides a ``clean'' beam with rear
spillover, ohmic loss, and blockage efficiencies near unity; therefore, the measured antenna temperature is roughly the same as the atmosphere-corrected antenna temperature, or $T_A \simeq T_A^*$.
For a point source, one can convert antenna temperature in Kelvin to flux density in \jy\ using 2.0\,\K\,Jy$^{-1}$ \citep{ghigo01}\footnote{see also\\ https://www.gb.nrao.edu/GBT/DA/gbtidl/gbtidl\_calibration.pdf}.  We can also convert the antenna temperature into the physical units of emission measure, ${\rm EM} = \int n_e^2 dl$, (see Appendix~\ref{sec:rrls}). Assuming an electron temperature of 8000\,\K,
a line width $\Delta V = 25\,\kms$, and the mean GDIGS \hna\ frequency of 5.7578\,\ghz\ (see Equation~\ref{eq:t_l}):
\begin{equation}
    \frac{{\rm EM}}{{\rm pc\,cm}^{-6}} = 176\times10^2\,\frac{T_{\rm L}}{\rm mK}\,,
    \label{eq:em}
\end{equation}
where $T_{\rm L}$ is the line intensity.
Given the range of measured values for $T_{\rm e}$ and $\Delta V$, we estimate that the relationship in Equation~\ref{eq:em} has $\sim 100\%$ uncertainty.
In terms of the integrated intensity, using Equation~\ref{eq:wrrl}, the conversion for \hna\ lines assuming an electron temperature of 8000\,\K is:
\begin{equation}
    \frac{{\rm EM}}{{\rm pc\,cm}^{-6}} = 6.39\,\frac{W_{\rm RRL}}{\rm mK\,\kms}\,.
    \label{eq:em2}
\end{equation}
Based on the range of values for \te in \hii\ regions and the DIG, this expression has $\sim\!50\%$ uncertainties.

%

\subsection{Observational Configuration}
The Versatile GBT Astronomical Spectrometer (VEGAS) can simultaneously tune to 64 spectral windows at two orthogonal polarizations (XX and YY).  Of these 64 tunings, we observe 22 \hna\ lines from $n = 95$ to 117 (excluding H113$\alpha$, which is compromised by the nearby H142$\beta$ line).   
We observe 25 \hnb\ lines from H120$\beta$ to H146$\beta$ (excluding H142$\beta$) and 8 \hng\ lines from H138$\gamma$ to H147$\gamma$ (excluding H142$\gamma$ and H145$\gamma$). We also observe transitions of H$_2$CO, CH$_2$NH, HC$_5$N, CCS, C$_3$S, HC$_9$N$_2$, CH$_3$OH, HCN, H$_2^{13}$CO and CH$_3$OH$_3$ (see Table~\ref{tab:lines} for transitions).
The molecular line data will be discussed in subsequent publications.  
Each spectral window spans 23.4\,\mhz\ with a spectral resolution of 2.86\,\khz.  GDIGS also observes the total-power continuum intensity, which can be computed from the measured system temperatures.  We are investigating whether the GDIGS continuum is a reliable data product.


In this observational configuration, the baseline stability is poor and the system temperature is elevated below 4.7\,\ghz\ and above 7.3\,\ghz.  This reduces the number of usable \hna\ lines to 15 (H97$\alpha$--H111$\alpha$) and the number of usable \hnb\ lines to 18 (H121$\beta$--H139$\beta$, excluding H137$\beta$, which is compromised by the 3(1,2)$\rightarrow$3(1,3) transition of CH$_3$OH$_3$).  Over the range of usable transitions the velocity resolution varies from $0.117$ to $0.182\,\kms$ and the full velocity range per tuning varies from  $961.6$ to $1494\,\kms$.

We list the usable spectral lines in Table~\ref{tab:lines}, which has columns of the line name, the transition, the line rest frequency, and the mean system temperature in the gridded maps (see Section~\ref{sec:maps}).  For all RRLs, in addition to the hydrogen lines we also simultaneously observe in the same bandpasses the RRLs of helium and carbon, since they are shifted by $\sim\!-120$ and $\sim\,-150$\,\kms\ from that of hydrogen, respectively.  This configuration was first used by \citet{anderson18} 
and was also employed by \citet{luisi18} and \citet{luisi19}.

\subsection{Survey Strategy\label{sec:strategy}}
\begin{sidewaysfigure*}
    \vskip -4in 
    \includegraphics[width=8.954in]{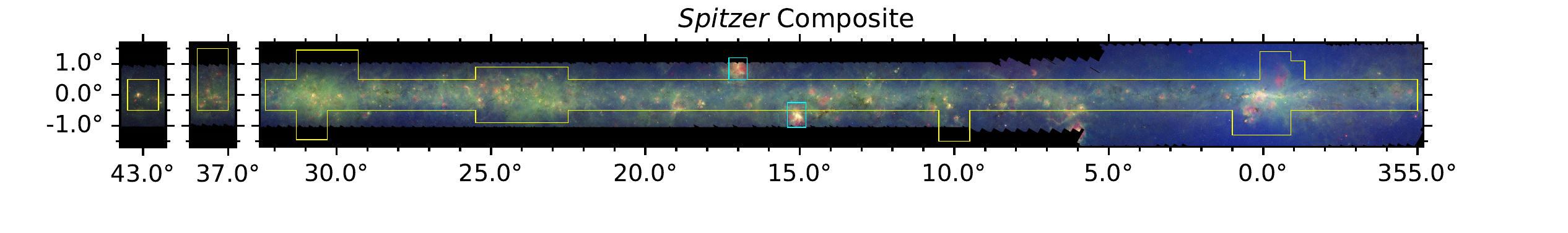}\\ \vskip -29pt
    \includegraphics[width=9in]{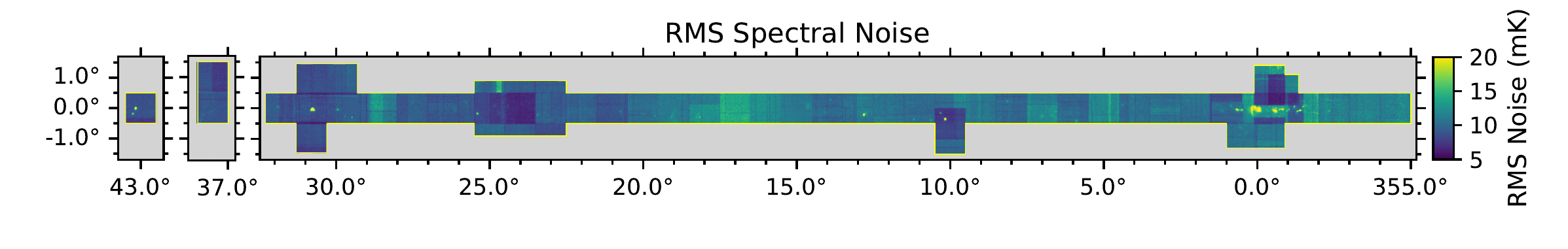}\\ \vskip -26pt
    \includegraphics[width=9in]{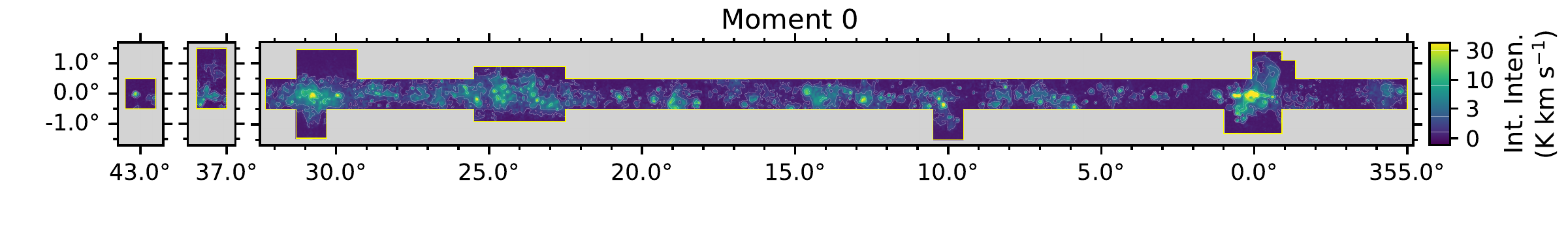}\\ \vskip -28pt
    \includegraphics[width=9in]{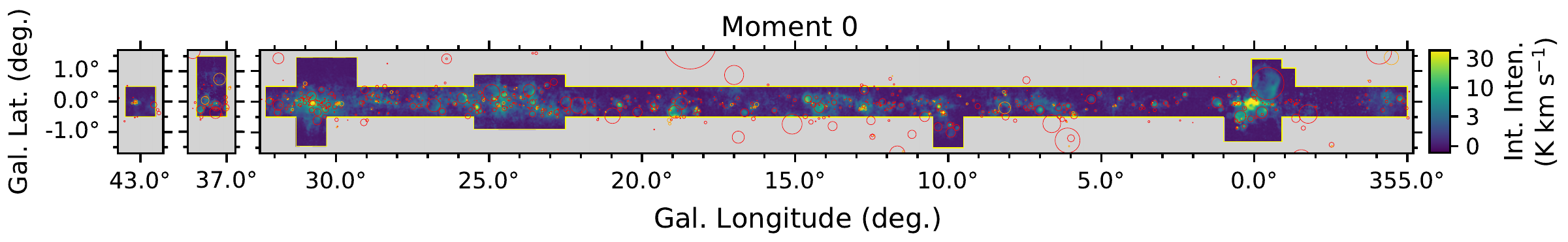}
    \caption{GDIGS areal coverage and intensity.  In all panels the yellow line shows the full GDIGS survey range. The nominal latitude range is $\absb < 0.5\degree$, with additional latitude coverage near W47 at $\ell\simeq 37\degree$, the giant \hii\ region W43 \citep{luisi20}, the Galactic center, $\ell \simeq 10\degree$, and $\ell \simeq 24\degree$.  {\it Top:} {\it Spitzer} data, with MIPSGAL \citep{carey09} 24\,\micron\ data in red, GLIMPSE \citep{benjamin03, churchwell09} 8.0\,\micron\ data in green, and GLIMPSE 3.6\,\micron\ data in blue.  In general, the regions of stronger 8.0\,\micron\ emission indicate intense radiation fields \citep[][their Figure~11]{luisi17} and \hii\ regions can be identified by the red 24\,\micron\ emission surrounded by green 8.0\,\micron\ emission \citep{anderson14}. The cyan  boxes denote coverage of M16 and M17 that are not part of the current data release, but were observed in the GDIGS configuration as part of the {\it SOFIA} Legacy project ``FEEDBACK'' (see Appendix~ \ref{sec:website}). {\it Top Middle:} Spectral noise (rms) in the \hna\ GDIGS data.  We compute the rms per 30\arcsec\ pixel with 0.5\,\kms\ channels; performing this calculation for 5\,\kms\ channels should decrease the rms by a factor of $3.2$. {\it Top Bottom:} GDIGS \hna\ integrated intensity (Moment 0) data.  Contours are at 0.5, 2, 8, and 30\,\K\kms.  One can convert from integrated intensity to EM using Equation~\ref{eq:em2}. {\it Bottom:} Same as the above integrated intensity panel, but with known \hii\ regions from the {\it WISE} Catalog overlaid in red and ``group'' \hii\ regions that appear to be associated with known regions in orange.}
\label{fig:coverage}
\end{sidewaysfigure*}

GDIGS covers $\sim\!49$ square degrees, as shown in Figure~\ref{fig:coverage}.  We describe the GDIGS data in this figure in later subsections.  The nominal survey zone, spanning $\sim\,37$~square degrees, is
$32.3\degree>\ell >-5\degree$, $\absb<0.5\degree$. We also map two square degrees in the area surrounding W47 ($38.0\degree > \ell > 37.0\degree$, $1.5\degree > b > -0.5\degree$) and 1~square degree in the area surrounding W49 ($43.5\degree > \ell > 42.5\degree$, $\absb < 0.5\degree$).   Due to the extended emission in the Galactic center, we add $\sim\, 4$~square degrees of additional latitude coverage there.  Finally, we provide a combined 5~ square degrees of additional latitude coverage within the nominal longitude zone of the survey to map extended \hii\ regions and trace bright diffuse ionized gas above and below the midplane; these areas are located near W31 ($\ell\simeq 10\degree)$, $\ell =24\degree$, and W43 \citep[$\ell\simeq 30\degree$;][]{luisi20}.  We describe additional mapped areas not included in this data release in Appendix~\ref{sec:website}.

\startlongtable
\begin{deluxetable}{llcr}
\tablecaption{Usable RRL and Molecular Line Transitions\label{tab:lines}}
\tablehead{
\colhead{Line} & Transition & \colhead{Frequency} & $\langle T_{\rm sys}\rangle$\\
\colhead{} & \colhead{} & \colhead{(\!\ghz)} & \colhead{(\!\mK)}
}
\startdata
     H97$\alpha$ & $n=98 \rightarrow 97$ &  7.09541 & 43.8\\
     H98$\alpha$ & $n=99 \rightarrow 98$ &  6.88149 & 34.4\\
     H99$\alpha$ & $n=100 \rightarrow 99$ &  6.67607 & 30.7\\
     H100$\alpha$ & $n=101 \rightarrow 100$ &  6.47876 & 25.8\\
     H101$\alpha$ & $n=102 \rightarrow 101$ &  6.28914 & 25.5\\
     H102$\alpha$ & $n=103 \rightarrow 102$ &  6.10685 & 24.9\\
     H103$\alpha$ & $n=104 \rightarrow 103$ &  5.93154 & 24.5\\
     H104$\alpha$ & $n=105 \rightarrow 104$ &  5.76288 & 23.3\\
     H105$\alpha$ & $n=106 \rightarrow 105$ &  5.60055 & 23.3\\
     H106$\alpha$ & $n=107 \rightarrow 106$ &  5.44426 & 22.4\\
     H107$\alpha$ & $n=108 \rightarrow 107$ &  5.29373 & 24.8\\
     H108$\alpha$ & $n=109 \rightarrow 108$ &  5.14870 & 25.0\\
     H109$\alpha$ & $n=110 \rightarrow 109$ &  5.00892 & 24.3\\
     H110$\alpha$ & $n=111 \rightarrow 110$ &  4.87416 & 24.4\\
     H111$\alpha$ & $n=112 \rightarrow 111$ &  4.74418 & 24.9\\\hline
     H121$\beta$ & $ n=123 \rightarrow 121$ &  7.24398 & 90.4\\
     H122$\beta$ & $ n=124 \rightarrow 122$ &  7.06882 & 60.5\\
     H123$\beta$ & $ n=125 \rightarrow 123$ &  6.89905 & 33.5\\
     H124$\beta$ & $ n=126 \rightarrow 124$ &  6.73479 & 32.6\\
     H125$\beta$ & $ n=127 \rightarrow 125$ &  6.57570 & 26.8\\
     H126$\beta$ & $ n=128 \rightarrow 126$ &  6.42158 & 27.6\\
     H127$\beta$ & $ n=129 \rightarrow 127$ &  6.27223 & 25.8\\
     H128$\beta$ & $ n=130 \rightarrow 128$ &  6.12748 & 26.3\\
     H129$\beta$ & $ n=131 \rightarrow 129$ &  5.98714 & 25.2\\
     H130$\beta$ & $ n=132 \rightarrow 130$ &  5.85107 & 23.7\\
     H131$\beta$ & $ n=133 \rightarrow 131$ &  5.71909 & 23.3\\
     H132$\beta$ & $ n=134 \rightarrow 132$ &  5.59105 & 23.8\\
     H133$\beta$ & $ n=135 \rightarrow 133$ &  5.46680 & 22.5\\
     H134$\beta$ & $ n=136 \rightarrow 134$ &  5.34619 & 23.4\\
     H135$\beta$ & $ n=137 \rightarrow 135$ &  5.22913 & 21.0\\
     H136$\beta$ & $ n=138 \rightarrow 136$ &  5.11544 & 24.0\\
     H138$\beta$ & $ n=140 \rightarrow 138$ &  4.89778 & 24.7\\
     H139$\beta$ & $ n=141 \rightarrow 139$ &  4.79357 & 24.6\\\hline
     H138$\gamma$ & $n=141 \rightarrow 138$ &  7.26889 & 128.5\\
     H139$\gamma$ & $n=142 \rightarrow 139$ &  7.11476 & 59.4\\
     H140$\gamma$ & $n=143 \rightarrow 140$ &  6.96495 & 33.3\\
     H141$\gamma$ & $n=144 \rightarrow 141$ &  6.81933 & 33.9\\
     H143$\gamma$ & $n=146 \rightarrow 143$ &  6.54004 & 26.4\\
     H144$\gamma$ & $n=147 \rightarrow 144$ &  6.40609 & 25.4\\
     H146$\gamma$ & $n=149 \rightarrow 146$ &  6.14898 & 23.6\\
     H147$\gamma$ & $n=150 \rightarrow 147$ &  6.02558 & 25.1\\\hline
     H$_2^{13}$CO & $ 1(1,0) \rightarrow 1(1,1) $ & ~4.59309\tablenotemark{$\dagger$} & 25.9\\ 
     H$_2$CO & $ 1(1,0) \rightarrow 1(1,1) $ &  4.82966 & 24.5\\  
     CH$_3$OH$_3$ & $ 3(1,2) \rightarrow 3(1,3)$ & ~5.00533\tablenotemark{$\dagger$} & 24.3\\       
     CH$_2$NH & $ 1(1,0) \rightarrow 1(1,1) $ &  5.28981 & 24.6\\ 
     HC$_5$N & $ J=2 \rightarrow 1 $ &  5.32533 & 24.0\\ 
     CCS & $ N = 3 \rightarrow 2,\, J = 2$ &  5.40260 & 22.8\\ 
     C$_3$S & $ 1 \rightarrow 0 $ &  5.78076 & 23.4\\ 
     HC$_9$N$_2$ & $ 10 \rightarrow 9 $ &  5.81036 & 23.5\\ 
     CH$_3$OH & $ 5(1,5) \rightarrow 6(0,6) $ &  6.66852 & 30.4\\ 
     HCN & $ J= 5,\, l=1e \rightarrow 1f $ &  6.73191 & 32.5\\ 
\enddata
\tablenotetext{\dagger}{H$_2^{13}$CO and CH$_3$OH$_3$ share their spectral windows with H141$\beta$ centered at 4.59384\ghz and H137$\beta$ centered at 5.00502\ghz, respectively; we do not use the H141$\beta$ and H137$\beta$ lines in our analyses.}
\end{deluxetable}


For the majority of the survey area, the default field size is 1 square degree.  We take data in on-the-fly (OTF) mode, slewing at 54\arcsec\,s$^{-1}$, with rows spaced every 40\arcsec\ ($0.39-0.25$ beams for the usable frequency range), and record data every 0.38\,s, or 20\arcsec\ ($0.19-0.12$ beams).  The row spacing is greater than the Nyquist rate at the highest usable frequency of 7.3\,\ghz. We observe each field in four coverages, making two complete maps by scanning in Galactic longitude and two more by scanning in Galactic latitude.  This redundancy reduces the impact of temporal artifacts such as weather.  For each of the four coverages, the integration time per beam is $4.6-11$\,s and therefore the total integration time per beam is $19-45$\,s. 

We observe a reference position every 16 rows, or $\sim\,20$\,minutes, using integration times of 66.6\,s (the duration of one row). This strategy ensures that all data in the map are taken within $\sim\!10$\,minutes of observing a reference position.  The reference positions are nominally 3\degree above the plane at the Galactic longitude of the field center.  We verify that each reference position is free of RRL emission prior to beginning a map by performing pointed observations at each position.  The pointed observations use the same setup as the OTF maps, with on- and off-source integration times
of 6 minutes each. The off-source scans track the same azimuth and zenith angle path as the on-source scans such that they follow the same path on the sky.  The typical rms spectral noise in the pointed observations is $\sim\!1.5$\,\mK (0.75\,\mjyb) after averaging all \hna\ lines and smoothing to 1.86\,\kms\ velocity resolution.  If emission is detected, we adjust the reference location and repeat the pointed observations prior to beginning the OTF maps.  

\begin{figure}
    \centering
    \includegraphics[width=0.47\textwidth]{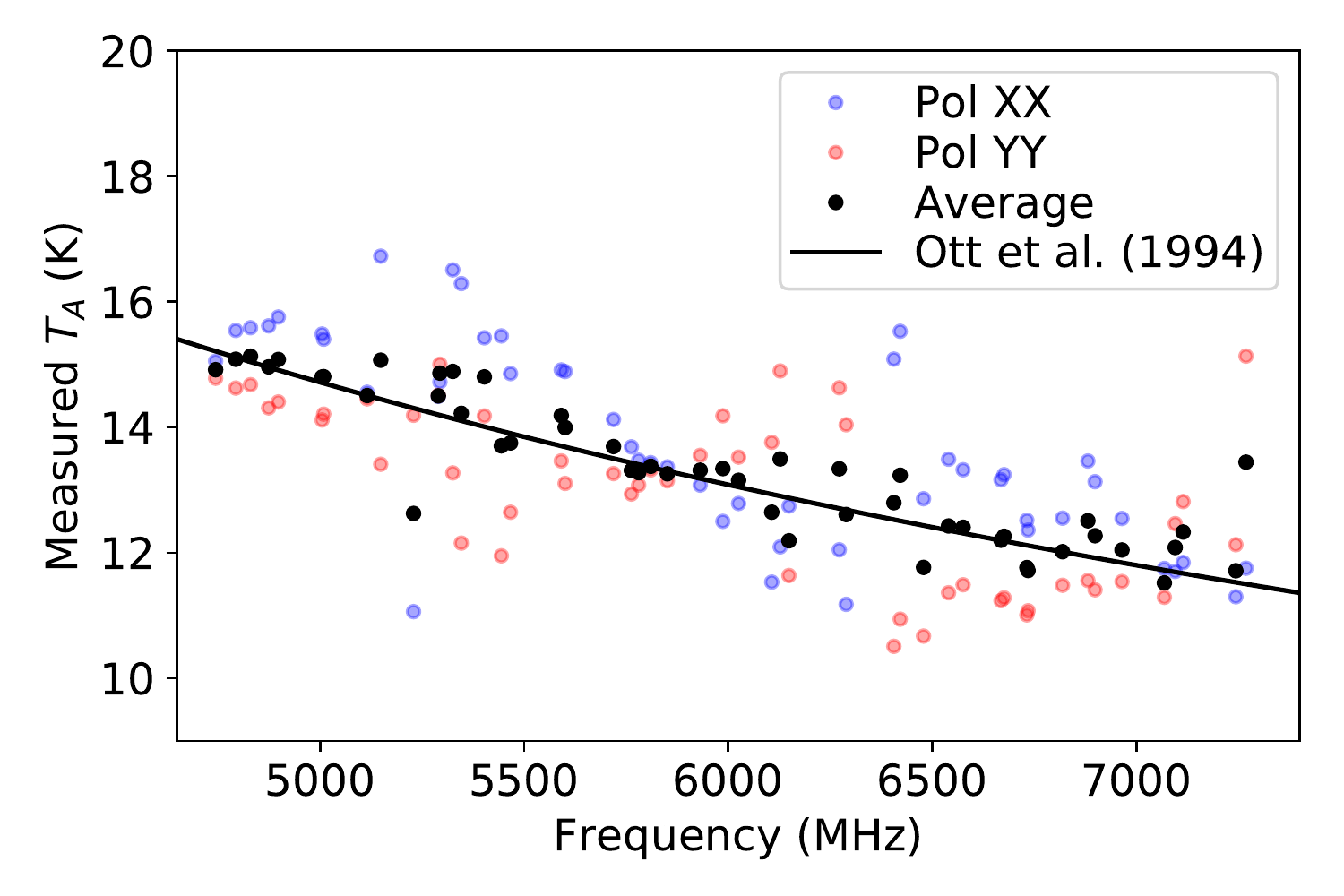}\\
    \includegraphics[width=0.47\textwidth]{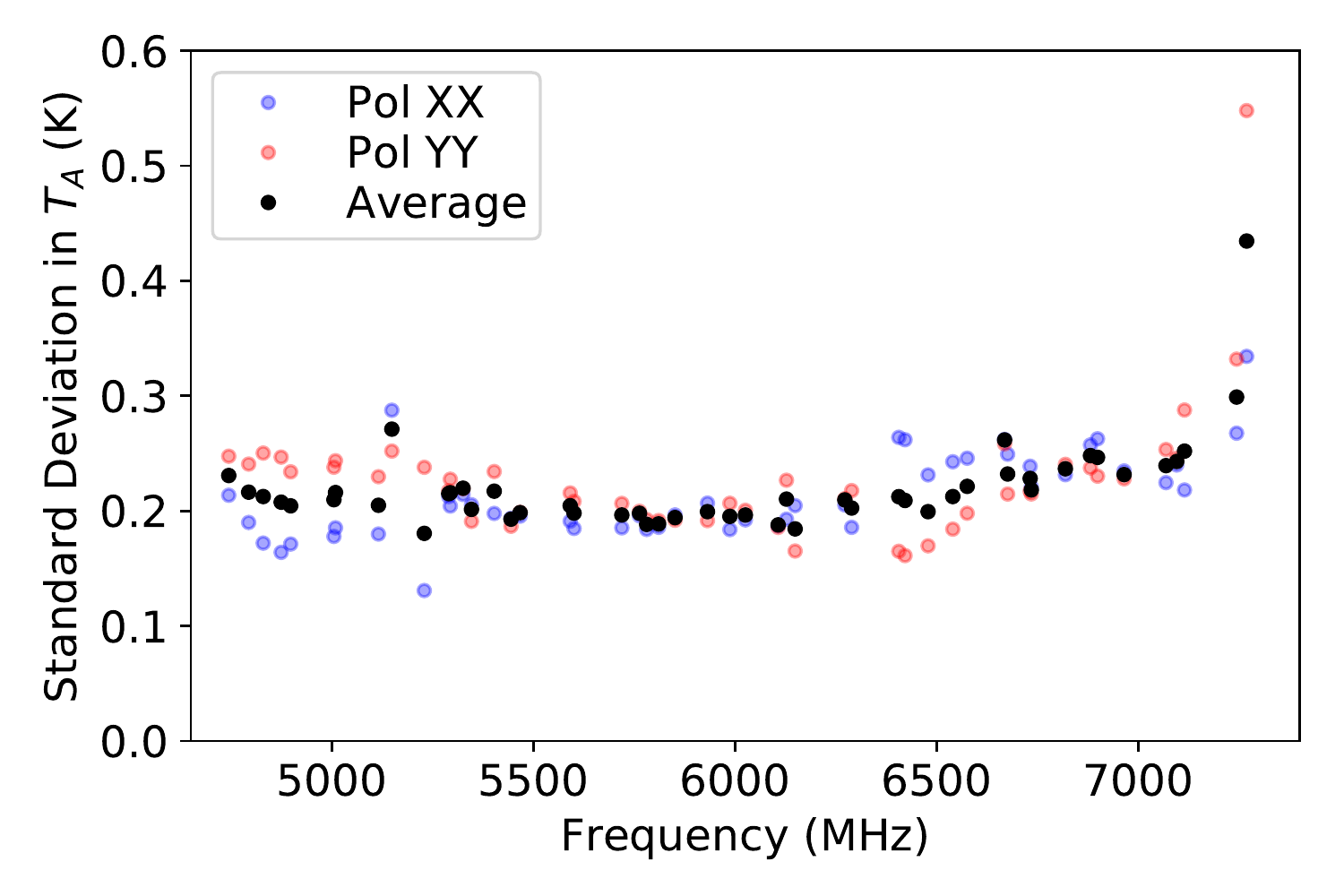}\\    
    \includegraphics[width=0.47\textwidth]{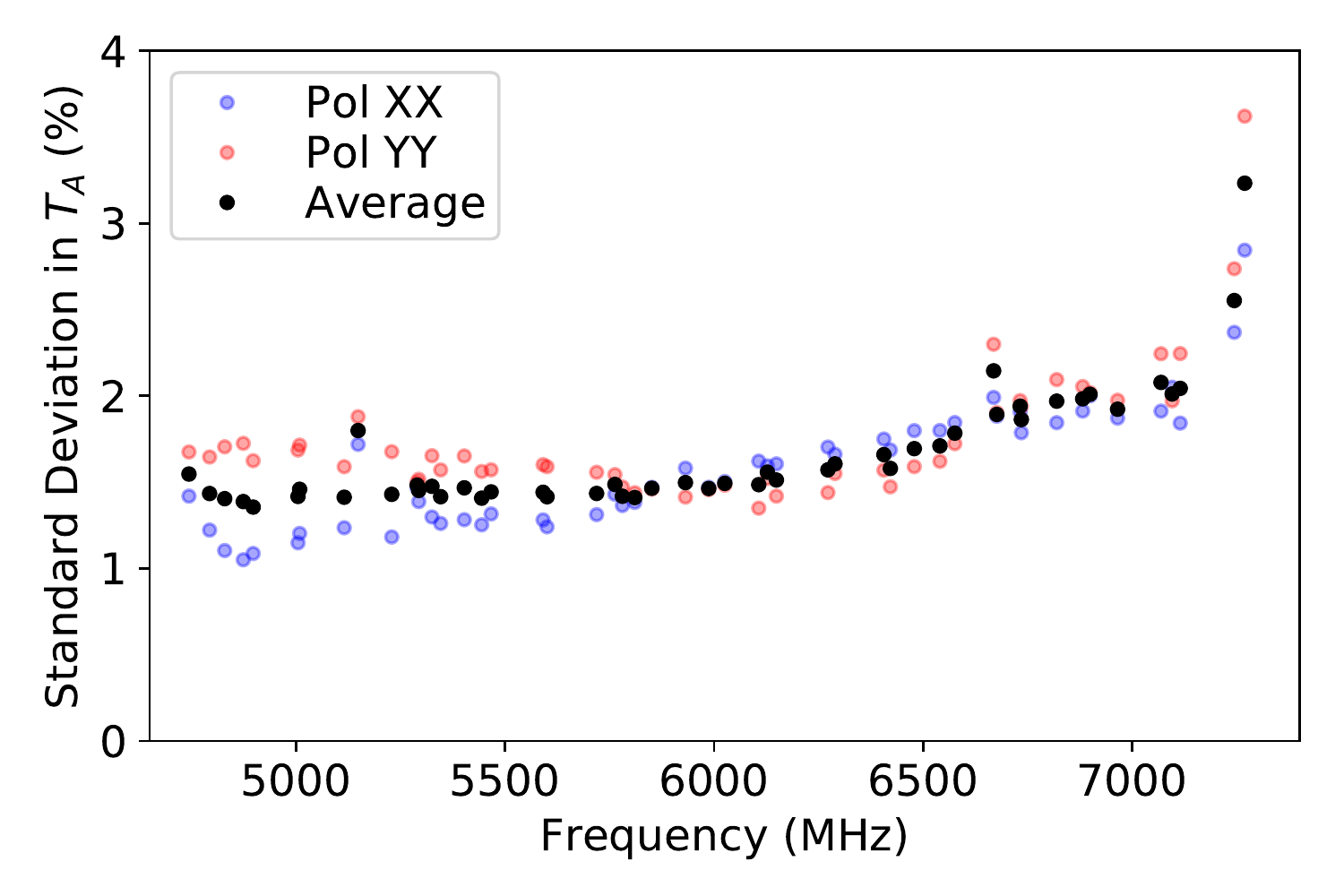}
    \caption{{\it Top:} Measured antenna temperatures of 3C286 for each spectral window and polarization prior to flux calibration. The antenna temperature values are averaged over all 14 observations of 3C286. The intensities of the individual polarization spectral windows show considerable deviations from the expected flux density of 3C286 \citep[][indicated by the black line]{ott94}, although much of this discrepancy disappears in the polarization-averaged data. {\it Middle:} Standard deviations in antenna temperature between all 14 maps of 3C286. {\it Bottom:} Same, but showing the relative standard deviations. The standard deviation averaged over all tunings is $1.7$\%.  \label{fig:cal}}
\end{figure}

\begin{figure}
    \centering
    \includegraphics[width=0.47\textwidth]{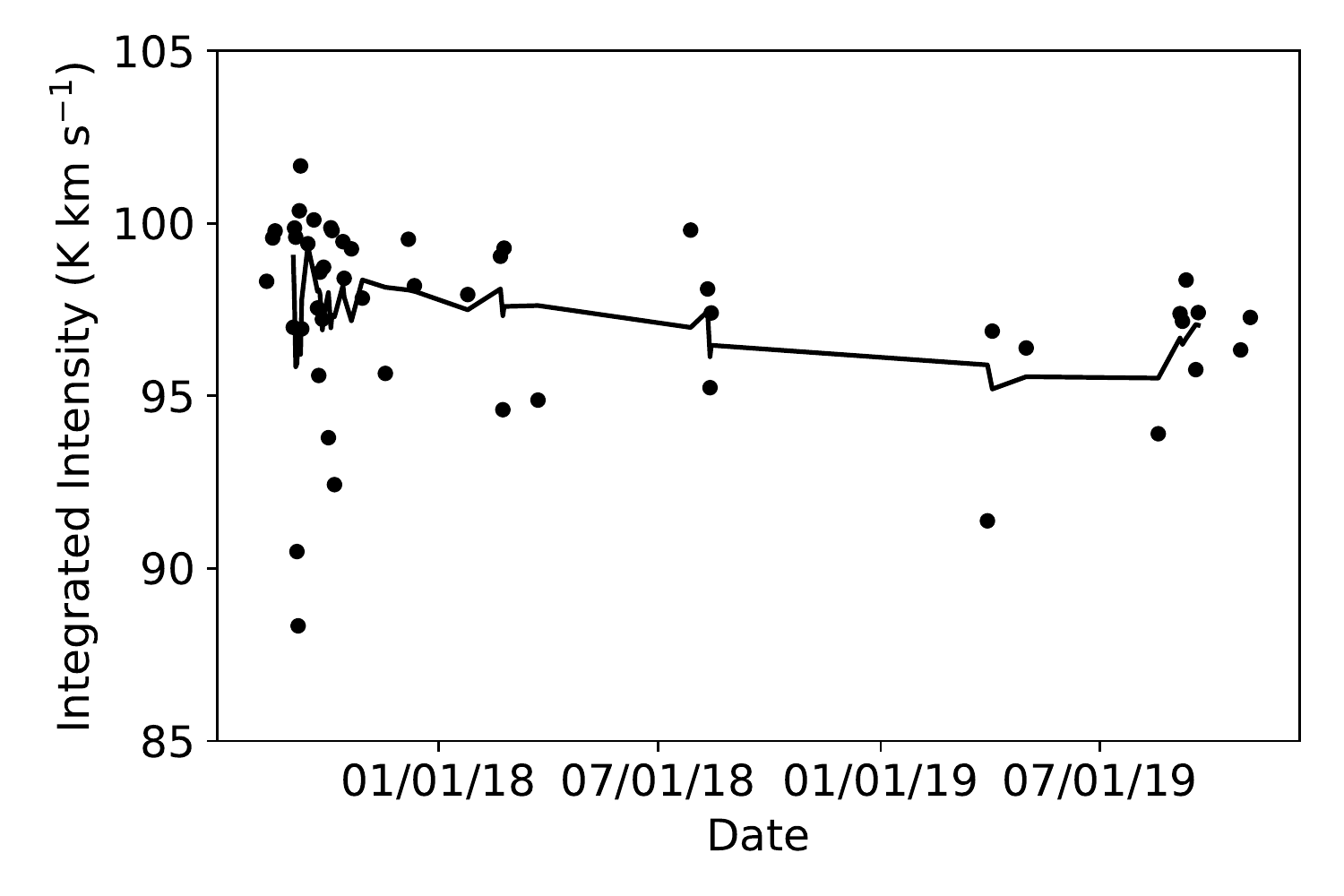}
    \caption{\hna\ integrated intensity of W43 as a function of observation date. The solid black line shows the moving average of the data with a subset size of 6. The standard deviation of all data is $3.6\%$ of the average value.  The apparent decrease in integrated intensity with time is not statistically significant ($p = 0.09$).\label{fig:w43}}
\end{figure}

\subsection{Calibration}
We calibrate the intensity scale of the data using two methods, but make no correction for elevation or weather.  We first calibrate the intensity scale
using noise diodes fired during data acquisition, a method that from experience has $\sim\!10\%$ uncertainties \citep{anderson11}. 

We also calibrate the intensity scale using the standard source 3C286, for which the flux density and spectral index are well-known.  We periodically (approximately every two months) map 3C286 using our same OTF mode and backend configuration; we make 14 maps in total.  As 3C286 is a nonthermal source, we do not measure spectral lines, only the total power continuum.  The use of 3C286 as a calibrator therefore assumes that the continuum and line responses can both be  calibrated using total power continuum observations. We determine the intensity of 3C286 in each map at each spectral window and polarization by measuring the amplitude of a two-dimensional Gaussian fit (which necessarily removes any spatial baseline).  

We average the measured 3C286 intensity of the 14 maps at each spectral window and polarization and compute the deviations from the expected intensity in \citet{ott94}.  These deviations are up to $\sim$10\% using the gain of 2.0\,\K\,Jy$^{-1}$.
(see Figure~\ref{fig:cal}, top panel).
We compute intensity corrections from the differences between the measured and expected values, which we then apply to all individual spectral window and polarization maps.
The standard deviation in the 3C286 intensities from the 14 individual spectral window and polarization maps has values of $1.5-2.0\%$ or $0.2\K$, increasing with frequency, which gives some indication of the variations due to weather, elevation, and electronics stability (see Figure~\ref{fig:cal}, middle and bottom panels).

Finally, to verify the stability of the spectral line intensity scale under the variety of observing conditions encountered, we periodically measure the average \hna\ emission from the giant \hii\ region W43 \citep[see][]{luisi20}.  
We observe W43 at the beginning of most observing sessions, using pointed observations with the same backend setup.  For each observation, we average all usable \hna\ transitions together, resample to a common velocity resolution of 1.86\,\kms, and fit a single Gaussian line profile. Figure~\ref{fig:w43} shows the integrated \hna\ intensities of all observations of W43. The distribution of observed integrated \hna\ intensities has a standard deviation of 2.8\%; the distribution of \hna\ intensities of W43 (not shown) has a standard deviation of 3.6\%.

The overall uncertainty in the intensity of the \hna\ data, computed as the combination of absolute calibration and temporal uncertainties, is $\sim\!5$\%.  We expect that the intensities of the average \hnb\ and \hng\ maps have similar uncertainties.  The pointed W43 data show that temporal effects lead to uncertainties in the average \hna\ data at the level of $\lsim\!4$\%.  From the standard deviation of the 3C286 data measured at each spectral window and polarization, the uncertainty is $\sim\!2$\%.  The temporal and absolute calibration effects are not entirely independent, as the mean value for the W43 integrated intensity is affected by the absolute calibration of the data. The estimated $\sim\!5\%$ uncertainty is in between the value obtained from independent (4.5\%) and dependent (6\%) absolute intensity and temporal uncertainties.

\subsection{RFI Removal}
Transitory radio frequency interference (RFI) can spoil entire spectra or cause abnormal spectral features in otherwise good spectra. 
We design two RFI mitigation techniques to remove spectra spoiled by RFI and to remove the strongest RFI signals from spectra that are otherwise usable.
For a given single-frequency and polarization spectrum, the first technique compares the rms of the line-free portions of the spectrum (approximately $-350$ to $-175\,\kms$ and $+175$ to $+350\,\kms$, using a narrower velocity range near the Galactic center) with the expected noise level from the radiometer equation that we calculate from the measured system temperature. We remove entire spectra that have rms values that are five times greater than that expected from the radiometer equation. 

In our second RFI removal operation, we first apply a median absolute deviation (MAD) filter to the data in each spectral window.  We remove data points that have absolute intensities $>6\times{\rm MAD}$, where MAD is computed in a sliding 101 channel window, and replace them with the median value over the same window. Since some low-intensity persistent RFI may still remain in the spectral window of the H104$\alpha$ line at high velocities, we apply the same MAD filter with a more aggressive $2\times{\rm MAD}$ threshold to the H104$\alpha$ data over the velocity range 125--200\kms. 

\begin{sidewaysfigure*}
    \vskip -4in 
    \includegraphics[width=8.954in]{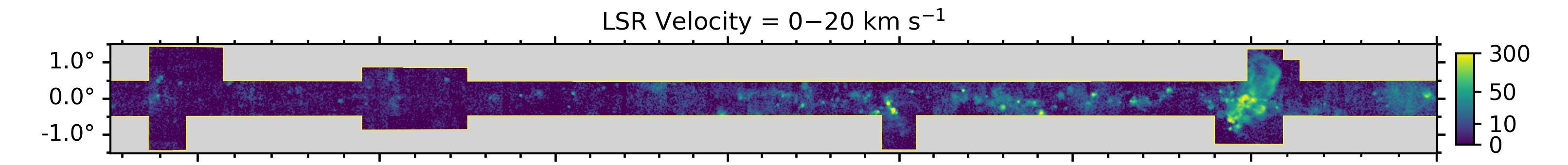}\\
    \includegraphics[width=8.954in]{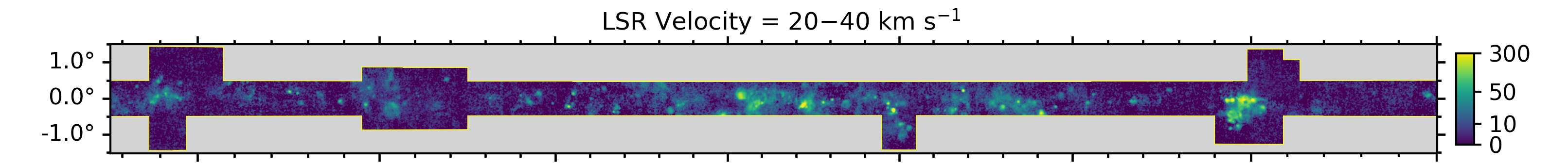}\\
    \includegraphics[width=8.954in]{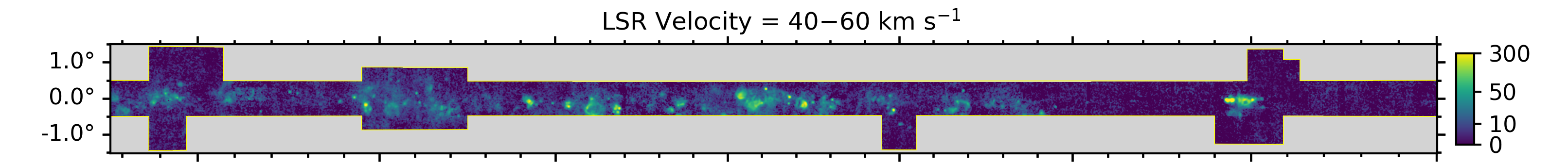}\\
    \includegraphics[width=8.954in]{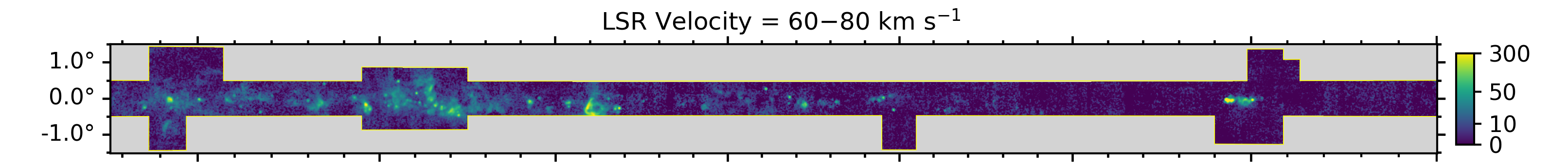}\\
    \includegraphics[width=8.954in]{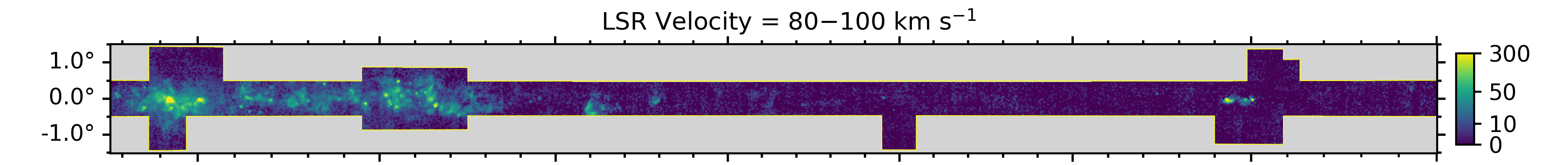}\\
    \includegraphics[width=8.954in]{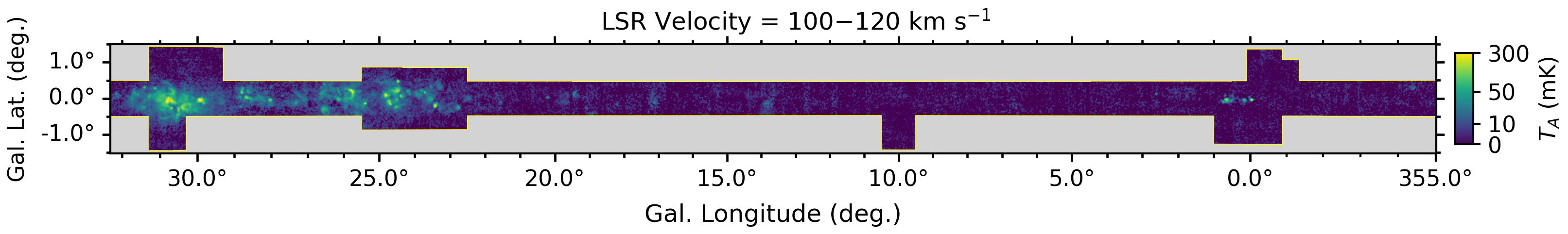}\\
\caption{GDIGS \hna\ emission in the large contiguous coverage area for local standard of rest (LSR) velocities between 0 and 120\,\kms, in 20\,\kms\ increments, smoothed to $2.5\arcmin$ resolution.  The intensity scale is inverse hyperbolic sine and ranges from 0 to 300\,\mK.  Due to Galactic rotation, the peak of the emission shifts from lower Galactic longitudes at lower velocities to higher Galactic longitudes at higher velocities.  RRL emission is especially broad in the Galactic center, and so emission in that region is visible in all panels.}
\label{fig:coverage_vel}
\end{sidewaysfigure*}

\begin{figure*}
    \centering
    \includegraphics[width=3.5in]{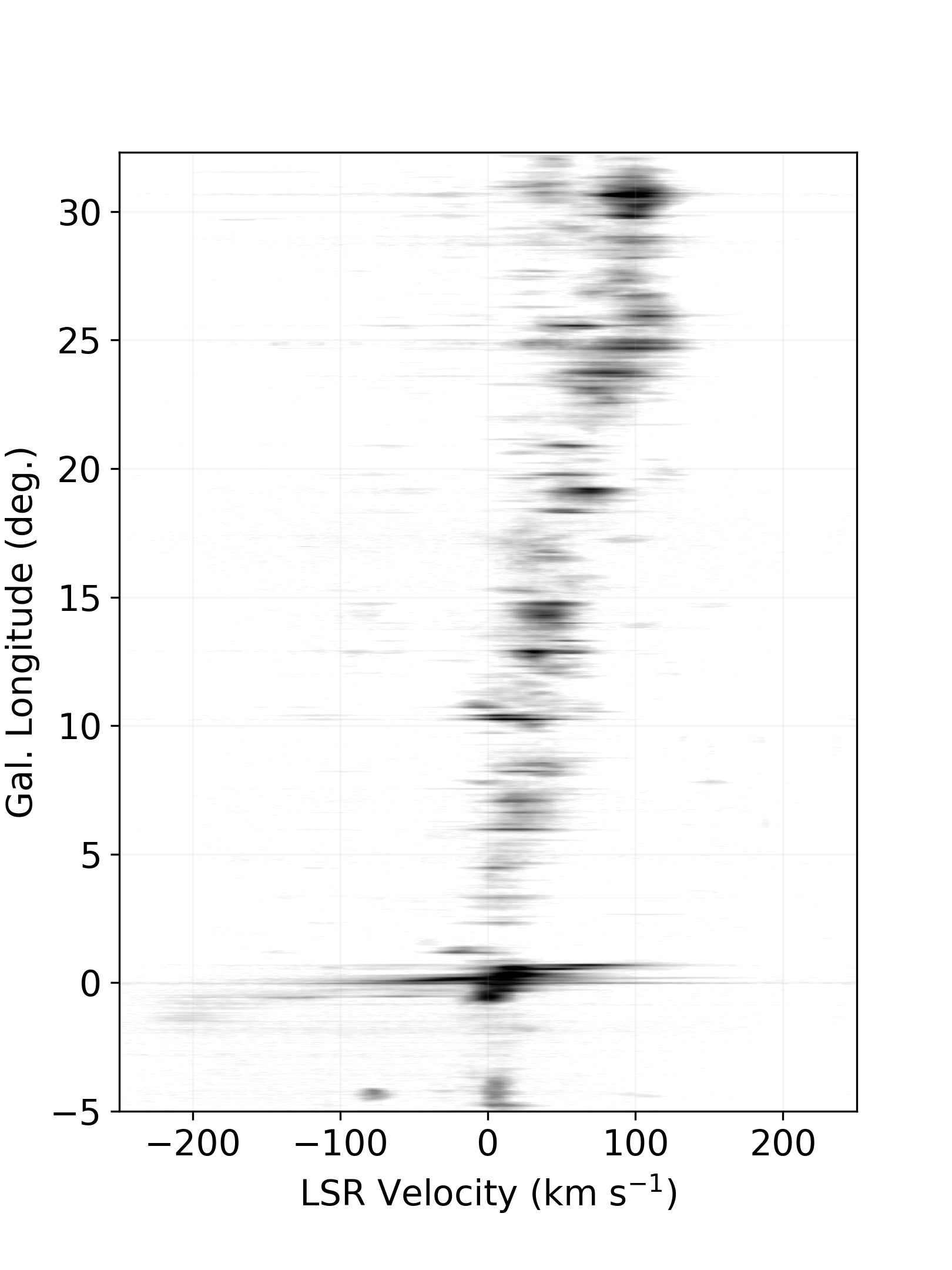}
    \includegraphics[width=3.5in]{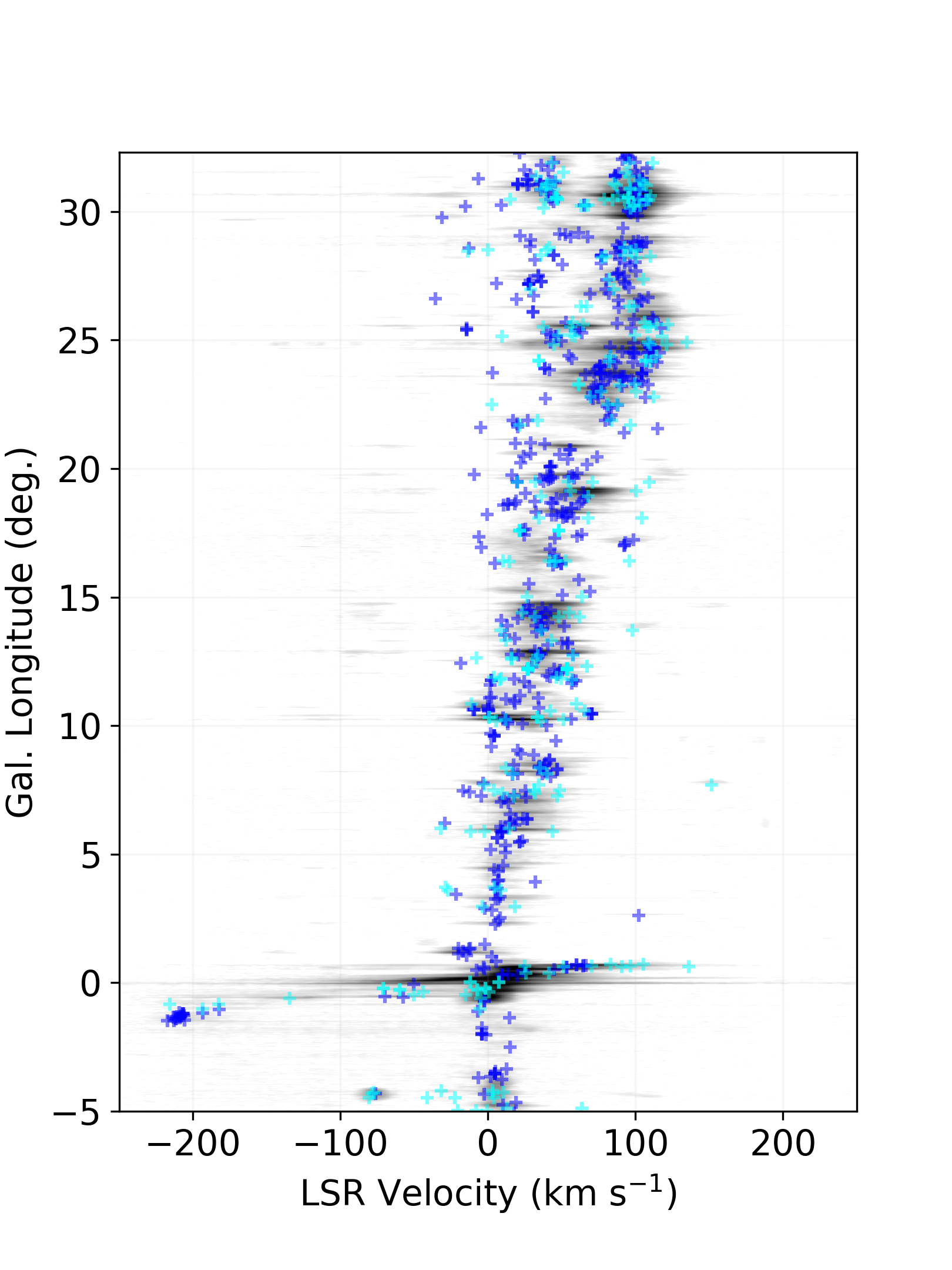}
    \caption{Longitude-velocity diagrams of the GDIGS \hna\ data, in grayscale, made by integrating over all latitudes.  The scale is linear from 0 to 10\,\K, and is the same in both panels.  Helium RRLs are evident, shifted by $\sim-120$\,\kms\ from that of hydrogen, at $\sim\!10\%$ the intensity.  The right panel overlays the positions of \hii\ regions from the {\it WISE} Catalog, with single-velocity \hii\ regions in blue and those with multiple detected RRL velocities in cyan.  Emission at velocities devoid of \hii\ regions is likely from the DIG.}
    \label{fig:lv}
\end{figure*}

\subsection{Map Making\label{sec:maps}}
Our goal is to produce average \hna, \hnb, and \hng\ maps with relatively uniform properties across the survey area.  The GBT spatial and spectral resolutions vary over the usable frequency range, so we must grid the data into cubes that have common spatial and spectral resolutions.

We use the {\it ``gbtgridder''}\footnote{https://github.com/GreenBankObservatory/gbtgridder} software to grid the RFI-removed spectra into single spectral window and polarization data cubes, with $30\arcsec$ square pixels.  There are $32-13$ pixels per beam for the usable frequency range.
The {\it gbtgridder} algorithm uses a Gaussian kernel, following the recommendations of \cite{mangum07}.
The native data cubes output from \textit{gbtgridder} have a third axis of frequency. 
We transform the third axis into units of velocity and re-sample to a uniform channel spacing of 0.5\,\kms\ using sinc interpolation (see T.~Wenger et al., 2021, in prep.) and a uniform velocity range of $-300\kms$ to $+300\kms$.
The {\it gbtgridder} algorithm automatically removes the median value of each spectrum;
we additionally subtract a 3rd-order polynomial baseline from each spectrum in the cubes determined from channels in the range $-300$ to $-200\kms$ and $+200$ to $+300\,\kms$.
If RFI is still present in the cubes after our mitigation techniques, we manually remove affected spectra and rerun \textit{gbtgridder}.

The advantage of observing a large number of transitions simultaneously is that we can average all lines at a given position to make one sensitive spectrum \citep{balser06}, a technique that is well-understood \citep{anderson11, liu13, alves15, luisi18}.  We smooth the individual maps to the resolution of the lowest frequency (4.74418\,\ghz\ for \hna, 2\arcmper65 full width at half maximum (FWHM) Gaussian kernel; 4.79457\,\ghz\ for \hnb, 2\arcmper62; and 6.02558\,\ghz\ for \hng, 2\arcmper09), and then average the individual maps pixel-by-pixel using a weighting factor of $T_{\rm sys}^{-2}$.
Finally, we average the two individual polarization data cubes.  This process creates average \hna, \hnb, and \hng\ data cubes at each spectral window.

Over the range $4.7$--$7.3\,\ghz$, we expect the RRL intensity to vary by $\sim\!60\%$ (see Equation~\ref{eq:t_l}).  For example, ignoring beam size effects, the intensity of the H111$\alpha$ line (in \K) at 4.74418\,\ghz\ will be 60\% higher than the intensity of the H97$\alpha$ line at 7.09541\,\ghz.  Taking into account the intensity variation with frequency and the $T_{\rm sys}^{-2}$ weighting we do when averaging, for average system temperatures at each spectral window the mean intensity-weighted frequency of the \hna\ lines is 
5.7578\,\ghz, 
which is near to the frequency of the 
H104$\alpha$ 
line.  
The mean intensity-weighted frequencies of the \hnb\ and \hng\ lines are 5.7959\,\ghz\ (in between H130$\beta$ and H131$\beta$) and 6.4528\,\ghz\ (nearest to H144$\gamma$), respectively. 

From these intensity-weighted frequencies, we can compute the expected intensity ratios of the $\alpha, \beta$, and $\gamma$ maps using Equation~\ref{eq:t_l}, assuming local thermodynamic equilibrium (LTE).  Ignoring beam size effects, the average \hnb\ maps should have intensity values 28\% that of the \hna\ maps; the average \hng\ maps should have intensity values 12\% that of the \hna\ maps, or 43\% that of the \hnb\ maps.

\subsection{Primary Data Products}
The primary GDIGS RRL data products discussed here are maps of average \hna, \hnb, and \hng\ emission with $30\arcsec$ pixels, $0.5\,\kms$ spectral resolution, and $\pm300\,\kms$ velocity coverage.  We also produce polarization-averaged maps of each individual \hna, \hnb, and \hng\ transition (H97$\alpha$ to H111$\alpha$; H121$\beta$ to H139$\beta$, excluding H137$\beta$; H138$\gamma$ to H147$\gamma$).  
These data, as well as data from the related projects (see Appendix~\ref{sec:website}) can be downloaded from our project web site\footnote{http://astro.phys.wvu.edu/gdigs/}.  For each data product, we produce 1~square degree maps, and also larger maps spanning $6\degree$ in longitude that overlap adjacent maps by $0.5\degree$ on the high- and low-longitude sides.  
If one \hna\ line has severe data quality issues, it is removed from the entire map.  If a given 1~square degree map has a compromised \hna\ line, all data from the line are excluded when creating the 1~square degree map and also when creating the corresponding map that spans $6\degree$ in longitude.  The sensitivity of the larger maps can thus be slightly lower than that of the 1 square degree maps, typically by a few percent.  For this reason, we recommend all studies of individual regions be performed with the 1 square degree maps. 

We show the integrated intensity (Moment~0) GDIGS \hna\ data in the bottom two panels of Figure~\ref{fig:coverage}.  To create this map, we 
regrid to a velocity resolution of 15\kms.  For each spaxel we then sum the emission from those velocity channels that have intensities greater than three times the rms spectral noise of that spaxel (see below).
One can convert from the integrated intensity value shown in the image to EM using Equation~\ref{eq:em2}.  Figure~\ref{fig:coverage_vel} provides an alternate view of the \hna\ data, with intensity in channels 20\,\kms\ wide.  
Galactic rotation is evident in Figure~\ref{fig:coverage_vel}, as the emission peaks at low Galactic longitudes at low velocities and higher Galactic longitudes at higher velocities.  Finally, in Figure~\ref{fig:lv} we show an \hna\ longitude-velocity diagram of the contiguous GDIGS zone, made by integrating over the complete latitude coverage.  This figure demonstrates that RRL emission is detected over the entire survey zone, and that the emission follows that of the \hii\ region distribution.

We compute the rms spectral noise 
over velocities $-300$ to $-150\,\kms$ and $+200$ to $+300$, and display the \hna\ noise map in Figure~\ref{fig:coverage}.  In the portion of the Galaxy observed by GDIGS, there should be almost no Galactic \hna\ emission in this velocity range. The one major exception is for Sgr~E \citep{liszt81, cram96}, for which we only use the positive velocity range when computing the rms.  
The noise is relatively uniform across the survey, but map edges and discrete locations of relatively high noise in the direction of bright continuum sources are apparent.

\begin{figure*}
    \centering
    \includegraphics[height=2.65in]{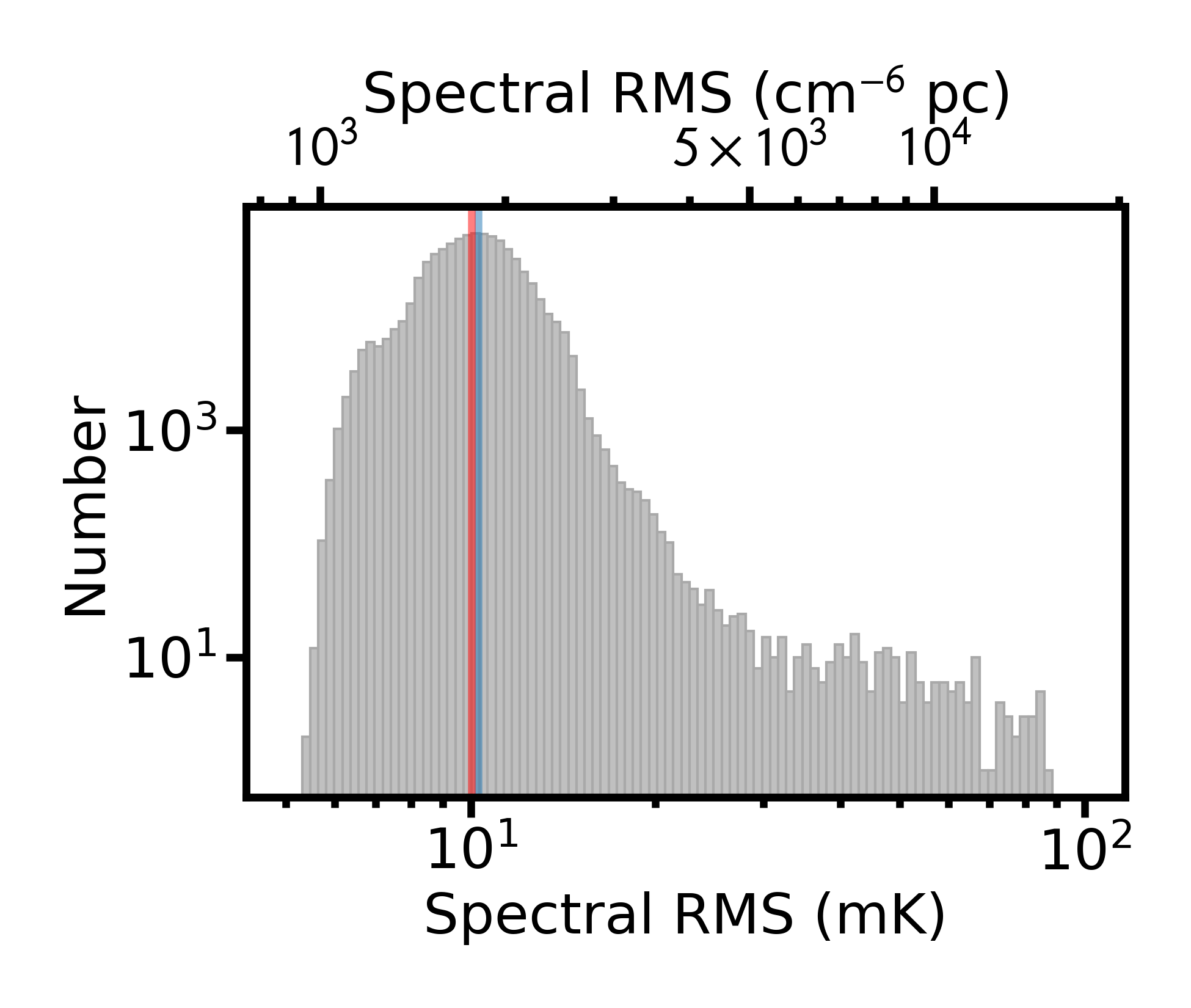}~~~
    \includegraphics[height=2.4in]{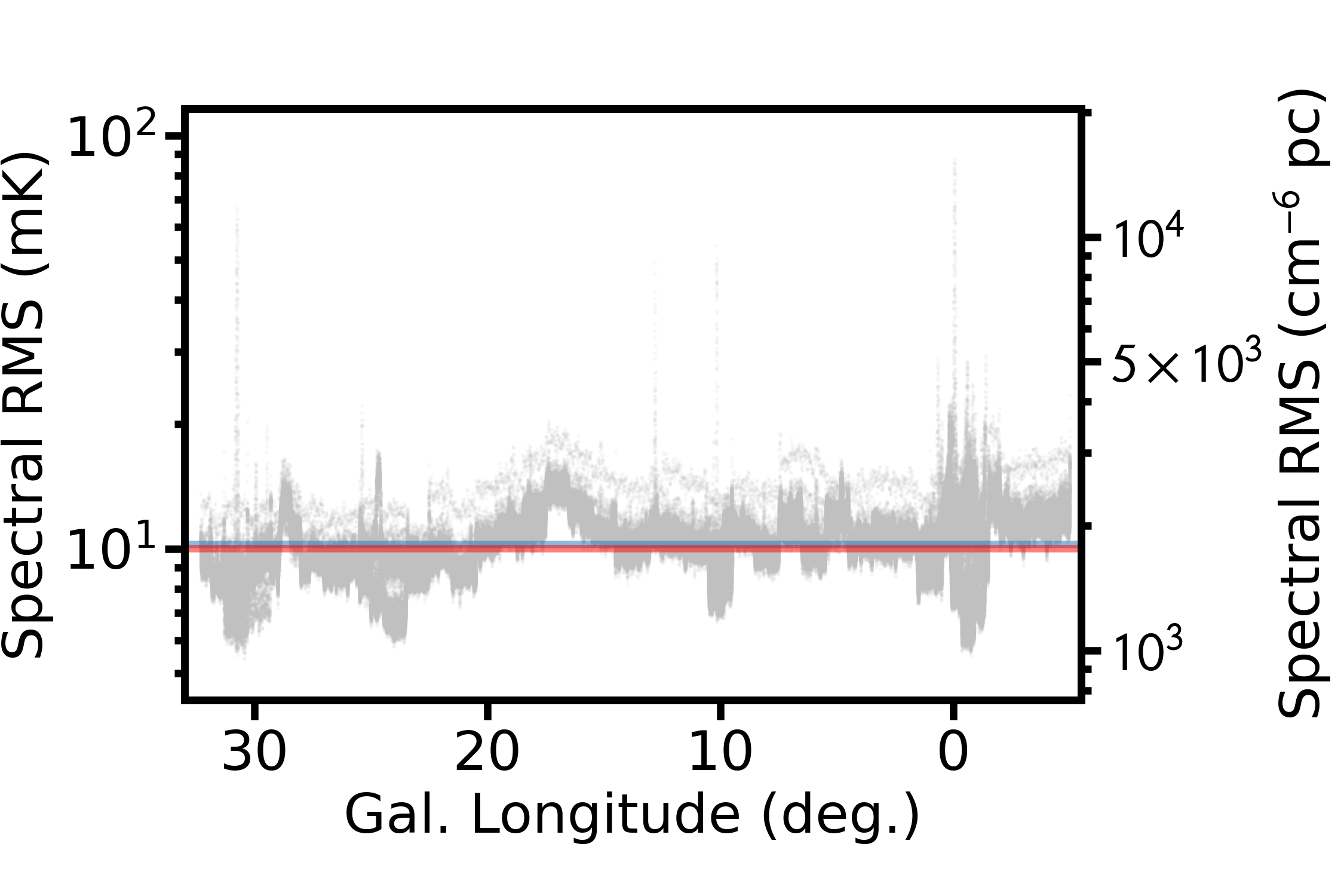}
    \caption{GDIGS \hna\ spectral RMS noise.  {\it Left:} The rms distribution over large contiguous regions of the survey (excluding areas around W47 and W49).  Both the mean (blue lines) and  mode (red lines) are near 10\,\mK (5\,m\!\jyb; 1100\,cm$^{-6}$\,pc).  {\it Right:} The rms noise as a function of Galactic longitude.  The noise increases slightly at lower Galactic longitudes as a result of low observing elevations.  The noise for 5\,\kms\ channels would decrease by a factor of 3.2 from the values shown.}
    \label{fig:noisehist}
\end{figure*}

For the entire survey, the mean and mode of the \hna\ rms spectral noise distribution as computed from the large cubes that span 6\degree\ in Galactic longitude
are $\sim\!10$\,mK (10.3\,mK and 10.0\,mK, respectively; see Figure~\ref{fig:noisehist}).  An rms of 10\,mK is equivalent to 5\,m\!\jyb or $\sim\!1800$\,cm$^{-6}$\,pc.  
Locations of high noise are found toward bright continuum sources, most notably at the location of W43 ($\ell\simeq 31\degree$) and Sgr\,A$^*$ ($\ell = 0\degree$).
Figure~\ref{fig:noisehist} shows that the noise also gradually increases at low Galactic longitudes, as a result of the low observing elevations.  The mean spectral noise in the \hnb\ maps is similar to that of the \hna\ maps (10.8\,mK), whereas due to the smaller number of RRLs, that of the \hng\ maps is $\sim 100\%$ higher (21.3\,mK).
Because RRLs are $\sim 25\,\kms$ wide, we can smooth to 5\,\kms\ resolution without loss of
information.  This smoothing should reduce the noise by a factor of 3.2.  Spatial smoothing can reduce the noise further.


We can relate the 10.3\,mK rms noise level to the properties of ionized gas using Equation~\ref{eq:em}.  
From Equation~1 of \citet{lenz92}, a $3\sigma$ detection requires a line intensity of $6.2$\,mK, for a line width of 25\,\kms\ and an rms noise level of 10.3\,mK.  Using Equation~\ref{eq:em2}, this corresponds to EM\,\,$\simeq 1100$\,cm$^{-6}$\,pc.
GDIGS is therefore sensitive to the \hna\ RRL emission from \hii\ regions with path lengths of 1\,pc and mean densities of $\langle n_e \rangle \gtrsim 30\,\percc$ (using Equation~\ref{eq:em}).
If the \hii\ region path length is 10\,pc, GDIGS can detect plasma with $\langle n_e \rangle \gtrsim 10\,\percc$.  For the DIG, if we assume a 1\,\kpc\ path length, GDIGS can detect plasma with $\langle n_e \rangle \gtrsim 1\,\percc$.  All values assume line widths of 25\,\kms\ and electron temperatures of 8000\,\K; the EM value that GDIGS is sensitive to will increase for narrower lines and/or higher electron temperatures (and hence so too will the values of $\langle n_e \rangle$).

\section{Enhanced Data Products}
Here, we describe and briefly characterize two enhanced data products provided for the community: maps resulting from automatic Gaussian decomposition and ``DIG-only'' data cubes.  The science enabled by these data products will be the subject of future papers in this series.

\subsection{Automatic Gaussian Decomposition}
\begin{figure*}
    \centering
    \includegraphics[width=3.5in]{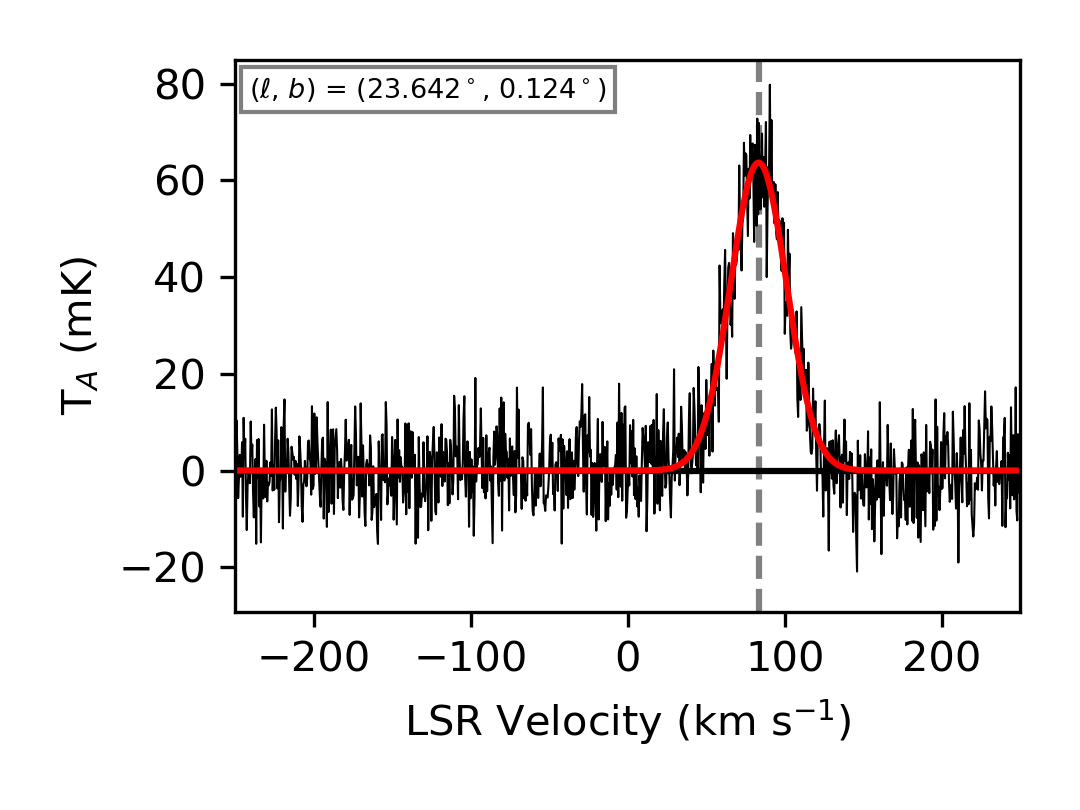}
    \includegraphics[width=3.5in]{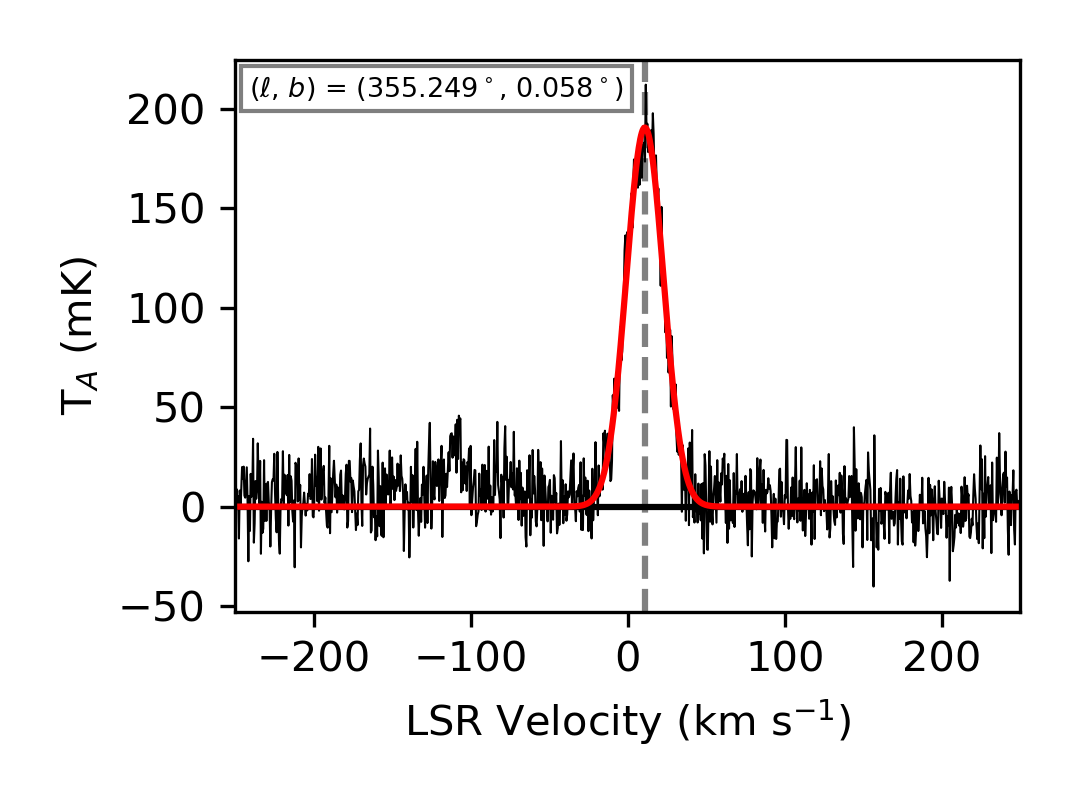}    
    \caption{Example \hna\ GDIGS spectra toward a position devoid of known \hii\ regions, \lb\ = (22.642\degree,0.124\degree) (left), and toward the \hii\ region G355.243+00.096, \lb\ = (355.249\degree, 0.058\degree) (right).  Each spectrum is from a single GDIGS spaxel. The automatic Gaussian decompositions are shown in red and the fitted central velocities are indicated with dashed vertical lines.  The He line is detected in the right panel at $\sim\!-120\,\kms$, but is excluded from the fits.}
    \label{fig:spectra}
\end{figure*}

A Gaussian decomposition of the GDIGS data allows us to study the RRL properties across the survey area.
We fit Gaussian profiles to the \hna\ GDIGS data using the ``GaussPy+'' algorithm \citep{riener19}.   This algorithm decomposes the spectral data into one or more Gaussian components for each spaxel that has a detection above the 3-sigma level, where the rms is evaluated independently for each spaxel.  We restrict the fits to have FWHM values above $5\,\kms$, which removes spurious fits to any residual RFI while retaining real signals \citep[see][their Figure~8]{anderson11}.
We also filter out any He RRL decompositions by removing fits whose peak velocities are offset $-130$ to $-110\,\kms$ from the peak velocities of brighter components at the same spaxels. 
We show representative \hna\ spectra and their automatic Gaussian fits in Figure~\ref{fig:spectra}.

\begin{figure*}
    \centering
    \includegraphics[width=6.5in]{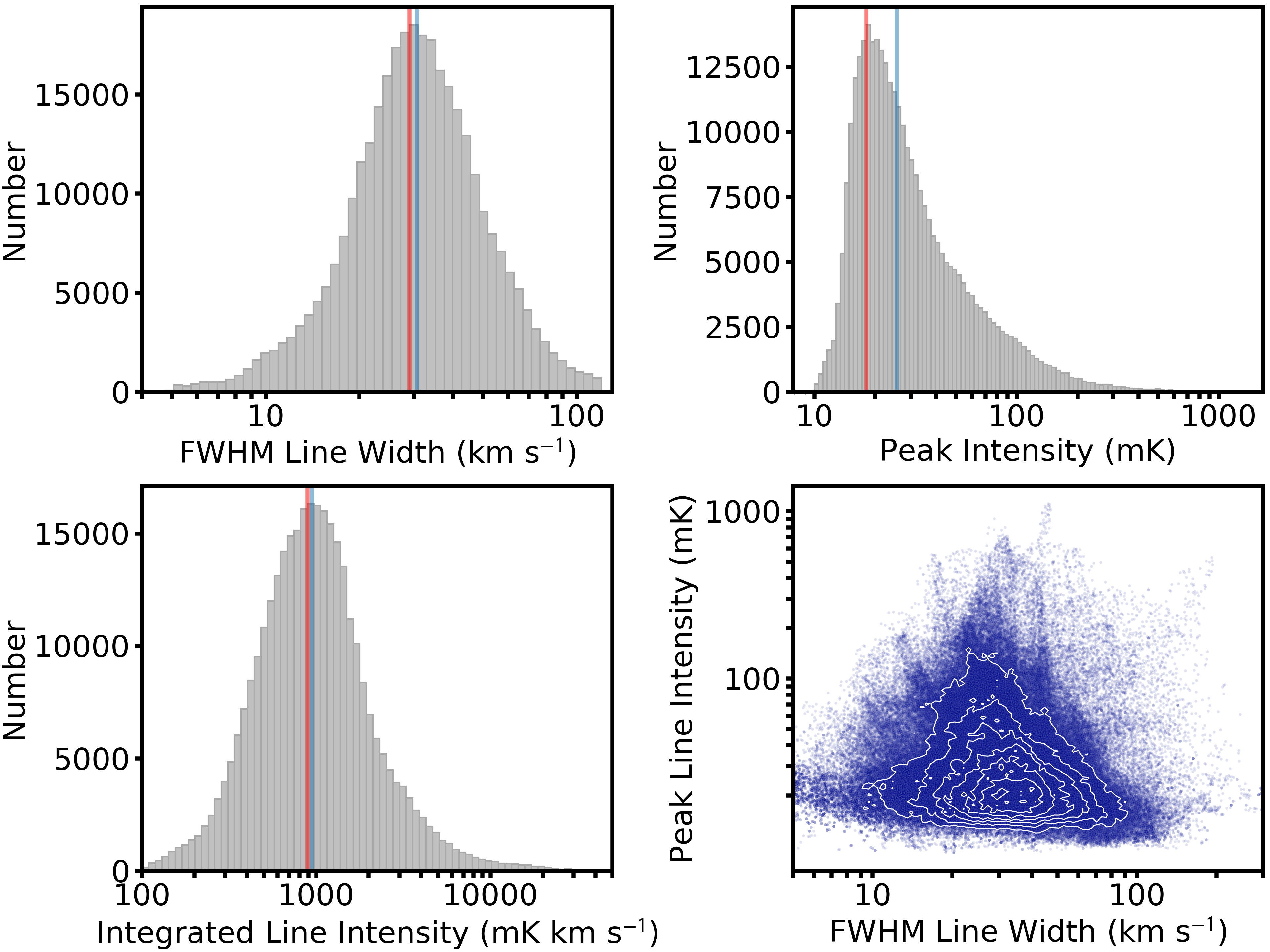}
\caption{Fit parameters from the automatic Gaussian decomposition.  {\it Top left}: FWHM line width distribution. The median is 30.6\,\kms\ and the mode is 29.0\,\kms. Here, and in the other histograms of this figure, the median values are indicated by blue vertical lines and the mode by red vertical lines. {\it Top right}: Peak intensity distribution. The median is 26\,\mK\ and the mode  is 18\,\mK.  {\it Bottom left}: Integrated line intensity distribution.  The median is 940\,\mK\kms\ and the mode is 890\,\mK\kms.  {\it Bottom right}: Peak line intensity versus FWHM line width.  There is no obvious correlation between the two parameters.
}
    \label{fig:gaussfits}
\end{figure*}

In Figure~\ref{fig:gaussfits}, we show histograms of the FWHM line widths, peak line intensities, and integrated line intensities obtained from this Gaussian decomposition, as well as a scatter plot of line intensity versus line width.  In each histogram we indicate the median (in blue) and mode (in red) of the distribution. The FWHM and integrated line intensity distributions are approximately log-normal, while the peak intensity has a high-intensity tail.  Although FWHM values $\gtrsim60\,\kms$ are not seen towards discrete \hii\ regions \citep[e.g.,][]{anderson11}, we believe the highest FWHM values in Figure~\ref{fig:gaussfits} are real, as they are mainly found towards the Galactic center where we expect broad lines due to the gas motions \citep[e.g.,][]{lockman73, geballe87}.
The FWHM versus intensity scatter plot shows no strong relationship between the quantities, but the brightest spectra have FWHM values $>20\kms$ and the narrowest lines are likely to be of low intensity.

\begin{figure*}
    {\myfont \large ~~~~~~~~~~~~~~~~~~~~~~~~GDIGS Hn$\alpha$~~~~~~~~~~~~~~~~~~~~~~~~~~~~~~~~~~~~~~~Reconstructed GDIGS Hn$\alpha$}\\
    \centering
    \includegraphics[trim=0 0 0 55, clip, width=3.2in]{LV_diagram_gdigs_rotated_noHII_newpipe_v5.png} 
    \includegraphics[trim=0 0 0 55, clip, width=3.2in]{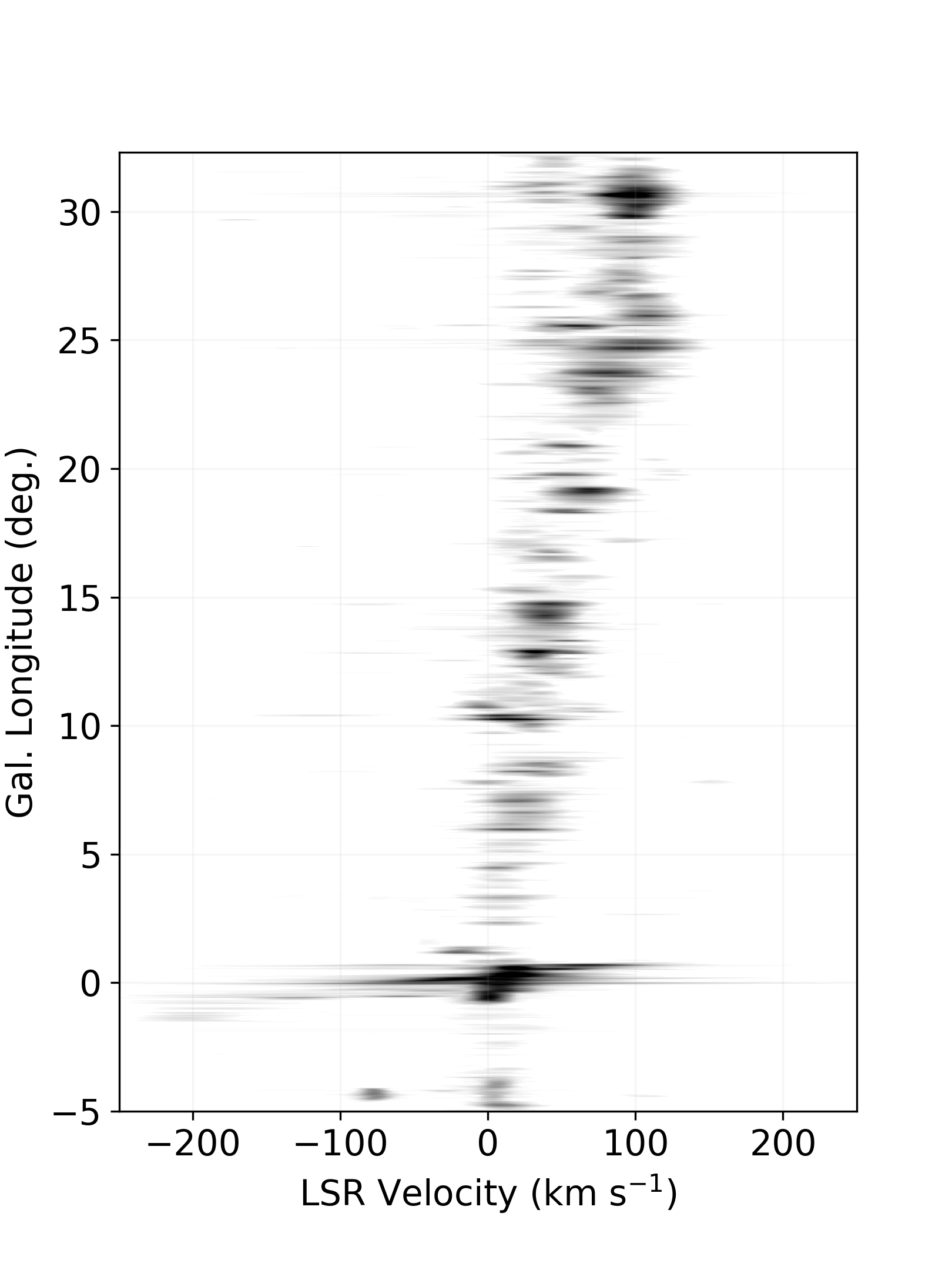}
    {\myfont \large ~~~~~~~~~~~~~~~~~~~~~~~~~~~~~~~~~~~~~~DIG-Only~~~~~~~~~~~~~~~~~~~~~~~~~~~~~~~~~~~~~~~~~~~~~~~~~~~~~~~~DIG-Only}\\
    \includegraphics[trim=0 0 0 58, clip,width=3.2in]{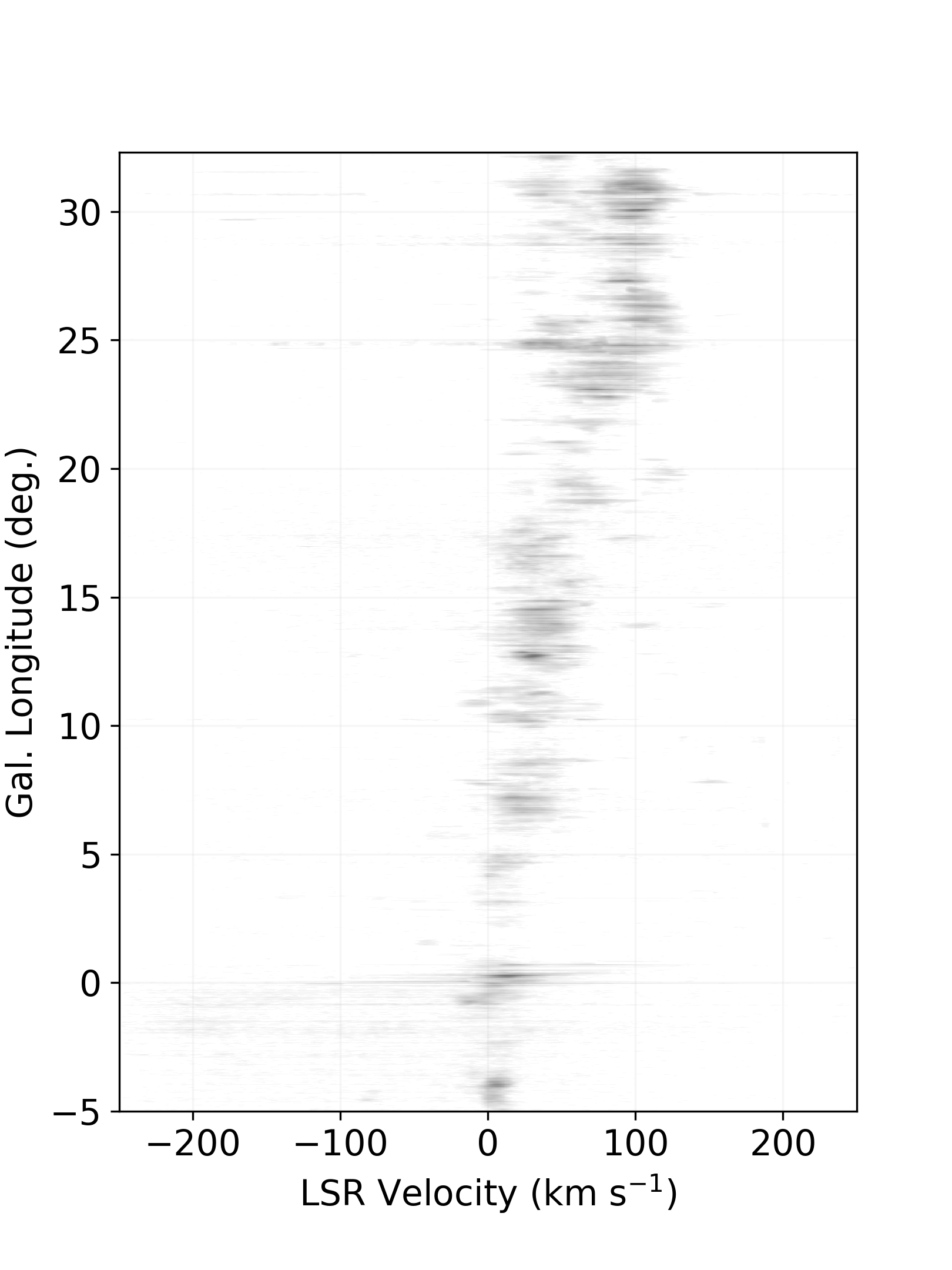}
    \includegraphics[trim=0 0 0 58, clip,width=3.2in]{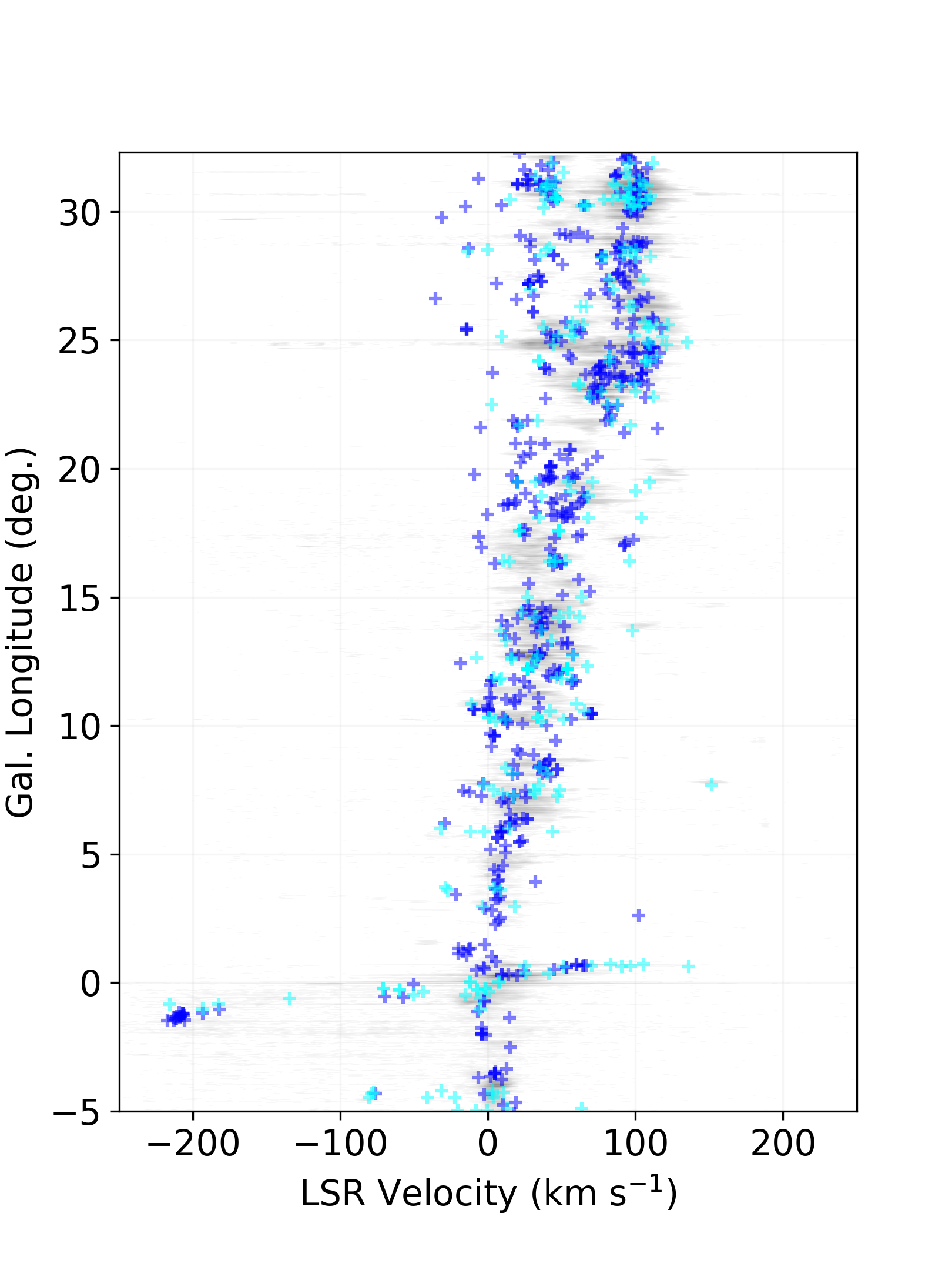}
    \caption{Longitude-velocity diagrams of the two enhanced data products, again integrated over all latitudes.  As in Figure~\ref{fig:lv}, the scale is linear ranging 0 to 10\,\K, and is the same in all panels.  The top left panel reproduces GDIGS \hna\ RRL data from Figure~\ref{fig:lv}.  The top right panel shows reconstructed GDIGS \hna\ RRL data produced from the automatic Gaussian decomposition.  The Helium RRLs, which are offset $\sim\!-120$\,\kms from that of hydrogen, have not been included in the automatic decomposition.  Comparison between these two top panels shows that the Gaussian decomposition recovers the bulk of GDIGS emission.
    The bottom two panels show DIG-only data, without \hii\ regions (left) and with (right), plotted in blue and cyan as in Figure~\ref{fig:lv}.  The DIG emission closely follows the distribution of \hii\ regions.}
    \label{fig:LV-synth}
\end{figure*}

Using results from the the automatic Gaussian decomposition, we also produce reconstructed data cubes.  For each spaxel with an automated fit, we fill the voxels of the reconstructed cubes using the fitted Gaussian parameters.  We then integrate the reconstructed cubes over all latitudes to create the longitude-velocity diagram in Figure~\ref{fig:LV-synth}. 
Figure~\ref{fig:LV-synth} shows that the Gaussian decomposition recovers nearly all of the GDIGS \hna\ emission, and that there is good agreement betweeen the decomposed GDIGS \hna\ data and the locations of discrete \hii\ regions.

\subsection{DIG-Only Data Cubes}
To facilitate analyses of the ionized gas components detected by GDIGS, we decompose the RRL intensity into emission from discrete \hii\ regions and diffuse ionized gas.  We can separate these components using data from the {\it WISE} Catalog of Galactic \hii\ Regions \citep[][hereafter {\it WISE} Catalog]{anderson14}.  The {\it WISE} Catalog is statistically complete for all \hii\ regions powered by O-stars (J.~Mascoop et al., 2021, submitted), and so should allow for a nearly complete census of the discrete \hii\ regions detected by GDIGS.

The {\it WISE} catalog has three main designations: known \hii\ regions with measured ionized gas spectroscopic velocities, candidate \hii\ regions that have the mid-infrared (MIR) morphology of \hii\ regions and radio continuum emission, and radio-quite candidates that have the MIR morphology but lack detected radio continuum emission.  A fourth category, ``group'' \hii\ regions, contains the \hii\ regions of larger complexes that were not measured in RRL emission individually.  The locations and angular sizes of \hii\ regions in the {\it WISE} Catalog were determined visually from MIR emission, including the emission of their PDRs.
Since PDRs are predominantly neutral, angular sizes from the {\it WISE} Catalog are typically larger than the sizes of the fully ionized \hii\ regions \citep{bihr16}.

We create ``DIG-only'' \hna\ data cubes devoid of emission from discrete \hii\ regions using the \hii\ region locations, angular sizes, line widths, and velocities from the {\it WISE} catalog.  For ``known'' \hii\ regions, we remove emission from all GDIGS \hna\ voxels at the measured \hii\ region \hna\ velocity within $\pm 1.5\times$ the measured \hna\ FWHM line width, for all spaxels that lie within the region (including pixels touching the \hii\ region boundary). We also remove potential helium RRLs using the same line width and location criteria, but with the velocity shifted by $-122\,\kms$ from that measured in hydrogen RRLs.  Assuming a Gaussian velocity profile, the intensity at $\pm 1.5\times$ the FWHM is $\sim\!0.2$\% of the peak intensity.  This cutoff ensures that any remaining emission from discrete \hii\ regions lies below our sensitivity threshold for a single velocity channel. For \hii\ region candidates, radio quiet candidates, and ``group'' \hii\ regions, we remove emission at all velocities, since we do not know the velocities of the regions. This may result in the loss of DIG emission, since diffuse gas potentially lies along the same line of sight as the \hii\ region candidates.
The DIG-only cubes are available on the GDIGS web site, and were first shown for the massive star forming region W43 in \citet{luisi20}, their Figure~9. The processing here produces data that are more sensitive by a factor of $\sim\!2$ over those of \citet{luisi20}, and therefore the data shown here supercede those of that paper.

We show the longitude-velocity diagram of these DIG-only data in Figure~\ref{fig:LV-synth}.  The DIG-only data lack the highest intensities (darker regions in the figures), which are due to discrete \hii\ regions.  The DIG emission shows excellent spatial agreement with the loci of \hii\ regions, indicating the close relationship between \hii\ regions and the DIG.  We will explore the nature of this relationship in a future paper.


\section{Summary}
GDIGS is a survey of ionized gas in the Galactic midplane.  It maps the C-band ($4-8$\,\ghz) RRL emission over $32.3\degree>\ell >-5\degree$, $\absb<0.5\degree$, with extended coverage above and below the plane in select fields and additionally covers W47 (around $\ell = 37.5\degree$) and W49 (around $\ell = 43\degree$).  The current data release concerns the \hna, \hnb, and \hng\ lines.

This paper characterized the GDIGS RRL data, and described enhanced data products that we provide to the community.  
The gridded GDIGS data have a pixel size of 30\arcsec\ and a channel width of 0.5\,\kms.  The \hna\ data have a spatial resolution of 2\arcmper65, whereas the \hnb\ and \hng\ data have spatial resolutions of 2\arcmper62 and 2\arcmper09, respectively.
The average spectral noise in the \hna\ data cubes is 10\,mK (5\,m\!\jyb), or $\sim\!1100$\,cm$^{-6}$\,pc assuming an electron temperature $\te = 8000\,\K$ and a line width of 25\,\kms.  

GDIGS gives us the clearest view yet of the large-scale distribution of ionized gas in the Milky Way.  Future papers in this series will examine the distribution of the DIG, the interplay between the DIG and discrete \hii\ regions, and the association between ionized gas and other components of the ISM.  These projects will shed new light on the properties of ionized gas in our Galaxy and its connection to high-mass star formation.

\begin{appendix}
\section{RRL Emission\label{sec:rrls}}
RRLs are produced 
in an ionized medium after an electron and an ion recombine.  The recombined atom (or ion if it was multiply ionized) can be in an excited state, and will emit RRLs as the electron cascades down the atomic levels toward the ground state. The hydrogen RRL photons are at frequencies
\begin{equation}
\nu_0 = R_{\rm M} c \left[\frac{1}{n^2} - \frac{1}{(n + \Delta n)^2}\right]\,,
\label{eq:nu}
\end{equation}
where $R_{\rm M}$ is the Rydberg factor for atoms of mass $M$, $c$ is the speed of light, $n$
is the principle quantum number of the final state, and $\Delta n$ is the change in principle quantum number.  
%
%
%

Transitions between adjacent energy levels, $\Delta n = 1$, are
referred to as $\alpha$ lines -- e.g., H89$\alpha$ is the
hydrogen $n = 90\rightarrow n = 89$ transition.  Transitions spanning
two levels, $\Delta n = 2$, are referred to as $\beta$ lines,
transition spanning three levels, $\Delta n = 3$, are $\gamma$
lines, etc.  \citep{lilley66}.  Hydrogen $\alpha$ transitions with principle
quantum numbers $n\gtrsim40$ are in the radio regime.  For $n\gtrsim40$ and
$\Delta n \ll n$, Equation \ref{eq:nu} becomes:
\begin{equation}
\nu_0 \approx \frac{2 (R_{\rm M} c) \Delta n}{n^3}\,.
\label{eq:nu_approx}
\end{equation}
The spacing between adjacent lines with the same $\Delta n$ is then
\begin{equation}
  \nu_0(n) - \nu_0(n-1) \approx -2 R_{\rm M} c \Delta n \frac{1}{n^3 - (n-1)^3} \approx -\frac{3\nu_0}{n}\,.
\end{equation}
Therefore, RRLs are spaced closer together at lower frequencies.

For \hna\ lines, the line center intensity of optically thin emission in LTE is \citep{gordon02,wenger19}: 
%
%
\begin{align}
\left(\frac{T_{\rm L}}{\K}\right) \simeq 3013 \left( \frac{T_{\rm e}}{\rm K} \right)^{-3/2} 
\left( \frac{\rm EM}{\rm cm^{-6}\,pc} \right) \left( \frac{\nu_0}{\rm GHz} \right)^{-1}
\left( \frac{\Delta V}{\kms} \right)^{-1} \frac{\Delta n}{n} f_{n,n+\Delta n}\,,
\label{eq:t_l_0}
\end{align}
where $\te$ is the electron temperature, EM is the emission measure, $\nu_0$ is the observed frequency, $\Delta V$ is the FWHM line width, and $f_{n,n+\Delta n}$ is the oscillator strength for transitions between levels $n$ and $n + \Delta n$.  \citet{menzel68} approximates the oscillator strength as
\begin{equation}
f_{n,n+\Delta n} \approx n M_{\Delta n} \left( 1 + 1.5 \frac{\Delta n}{n} \right)\,,
\label{eq:osc}
 \end{equation}
where $M_{\Delta n}$ has values of $M_{\Delta n} = 0.190775, 0.026332,$ and 0.0081056 for $\Delta n = 1,2,$ and 3, respectively \citep{menzel68}.
Substituting $\nu_0$ for $n$ using Equation~\ref{eq:nu_approx} and using the expression for the oscillator strength in Equation~\ref{eq:osc}, we find
\begin{align}
\left(\frac{T_{\rm L}}{\K}\right) \simeq 3013 \left( \frac{T_{\rm e}}{\rm K} \right)^{-3/2} 
\left( \frac{\rm EM}{\rm cm^{-6}\,pc} \right) \left( \frac{\nu_0}{\rm GHz} \right)^{-1} \left( \frac{\Delta V}{\kms} \right)^{-1} \Delta n M_{\Delta n} \left[ 1 + 0.00800 \Delta n^{2/3} \left(\frac{\nu_0}{\rm GHz} \right)^{1/3} \right]\,.
\end{align}
Since $1 + 0.00800 \Delta n^{2/3} \left( \nu_0 /\ghz \right)^{1/3} \simeq 1.01-1.03$ for the frequencies and values of $\Delta n$ used here, using an average value of 1.02 we find
\begin{align}
\left(\frac{T_{\rm L}}{\K}\right) \simeq 3076 \left( \frac{T_{\rm e}}{\rm K} \right)^{-3/2} 
\left( \frac{\rm EM}{\rm cm^{-6}\,pc} \right) \left( \frac{\nu_0}{\rm GHz} \right)^{-1} \left( \frac{\Delta V}{\kms} \right)^{-1} \Delta n M_{\Delta n}\,.
\label{eq:t_l}
\end{align}

Assuming the line width is not a function of frequency, for a given $\Delta n$ the measured intensity decreases approximately linearly with increasing frequency.  Since the flux density in Jy follows $S_\nu \propto T_{\rm L} \nu^2$, it increases approximately linearly with increasing frequency. For a Gaussian line profile, the RRL integrated intensity $W_{\rm RRL}$ is:
\begin{equation}
    \frac{W_{\rm RRL}}{\rm K\,\kms} = \frac{1}{2}\left(\frac{\pi}{\ln 2}\right)^{0.5} \frac{T_L}{\rm K} \frac{\Delta V}{\kms} \simeq 3274 \left( \frac{T_{\rm e}}{\rm K} \right)^{-3/2} 
\left( \frac{\rm EM}{\rm cm^{-6}\,pc} \right) \left( \frac{\nu_0}{\rm GHz} \right)^{-1}  \Delta n M_{\Delta n}\,.
 \label{eq:wrrl}
\end{equation}
These relations neglect beam size effects.


\section{GDIGS Web Site\label{sec:website}}
We provide a web site so users can download GDIGS data\footnote{http://astro.phys.wvu.edu/gdigs/}.  This site currently contains \hna, \hnb, and \hng\ data.  We will continue to expand this site as more GDIGS data are published.

The web site also contains data from related projects taken with the same GDIGS observing mode, calibration, and data reduction, albeit often with different mapping speeds. Table~\ref{tab:other} lists these projects and has columns of the field name,  field centroid, mapped area, and integration time per \hna\ $2\arcmper65$ beam.  As part of the {\it SOFIA} ``FEEDBACK'' Legacy project that maps \cii\ emission from \hii\ region complexes \citep{schneider20}, we observed fields around M17, M16, W40, DR21, and NGC7538.  We observed fields around S235 to aid the analysis of {\it SOFIA} \cii\ data in \citet{anderson19b} and around W51 to investigate the cause of high rotation measures found in this direction \citep{shanahan19}.  Finally, we observed fields around Cygnus~X and Orion as part of an ongoing effort to map the brightest star forming regions in the Galaxy.

\begin{deluxetable}{lrrrc}[!h]
\tablecaption{Other Observed Targets\label{tab:other}}
\tablehead{
\colhead{Field} &
\colhead{$\ell$} & 
\colhead{$b$} & 
\colhead{Area} &
\colhead{$t_{\rm int}$}\\
\colhead{} &
\colhead{(deg.)} &
\colhead{(deg.)} &
\colhead{(sq. deg.)} &
\colhead{(s)}
}
\startdata
M17 & $15.10\degree$ & $-0.70\degree$ & 0.42 & 45\\
M16 & $17.00\degree$ & $0.85\degree$ & 0.42 & 45\\
W40 & $28.75\degree$ & $3.45\degree$ & 0.36 & 45\\
W51 & $48.94\degree$ & $0.00\degree$ & 8.75 & 18\\ 
Cygnus~X & $79.74\degree$ & $0.88\degree$ & 64.00 & \phn 3.8\\
DR21 & $81.65\degree$ & $0.65\degree$ & 0.42 & 36\\
NGC7538 & $111.60\degree$ & $0.85\degree$ & 0.25 & 45\\
S235 & $173.60\degree$ & $2.70\degree$ & 0.56 & 45\\
Orion & $209.00\degree$ & $-19.50\degree$ & 3.20 & 45
\enddata
\end{deluxetable}

\end{appendix}

\begin{acknowledgments}
This work is supported by NSF grant AST1516021 to LDA.  
\nraoblurb\
The Green Bank Observatory is a facility of the National Science Foundation operated under cooperative agreement by Associated Universities, Inc.
We
thank the staff at the Green Bank Observatory for their hospitality and friendship during the observations and data
reduction. We thank West Virginia
University for its financial support of GBT operations, which
enabled some of the observations for this project. \end{acknowledgments}


\facility{Green Bank Telescope}

\software{AstroPy \citep{astropy2013, astropy2018}, GBTIDL\footnote{http://gbtidl.nrao.edu/} \citep{gbtidl}, \textit{gbtgridder}\footnote{https://github.com/GreenBankObservatory/gbtgridder}, GaussPy+ \citep{riener19}}

\bibliographystyle{aasjournal}
\bibliography{ref.bib}

\end{document}

%% file: preamble.tex
\usepackage{xspace}
\usepackage{graphicx}
\usepackage{natbib}
\usepackage{amsmath}
\usepackage{booktabs}
\usepackage{multirow}
\usepackage{afterpage}
\usepackage[counterclockwise, figuresleft]{rotating}
\usepackage[bottom]{footmisc}

\newcommand{\nraoblurb}{The National Radio Astronomy Observatory is
a facility of the National Science Foundation operated under cooperative
agreement by Associated Universities, Inc.}
\newcommand{\hide}[1]{}

\newcommand{\lsim}{\ensuremath{\,\lesssim\,}\xspace}
\newcommand{\gl}{\ensuremath{\ell}\xspace}
\newcommand{\gb}{\ensuremath{{\it b}}\xspace}
\newcommand{\absb}{\ensuremath{\vert\,\gb\,\vert}\xspace}

\newcommand{\lb}{\ensuremath{(\gl,\gb)}\xspace}

\newcommand{\kms}{\ensuremath{\,{\rm km\,s^{-1}}}\xspace}

\newcommand{\pc}{\ensuremath{\,{\rm pc}}\xspace}
\newcommand{\kpc}{\ensuremath{\,{\rm kpc}}\xspace}
\newcommand{\K}{\ensuremath{\,{\rm K}}\xspace}
\newcommand{\mK}{\ensuremath{\,{\rm mK}}\xspace}

\newcommand{\khz}{\ensuremath{\,{\rm kHz}}\xspace}
\newcommand{\mhz}{\ensuremath{\,{\rm MHz}}\xspace}
\newcommand{\ghz}{\ensuremath{\,{\rm GHz}}\xspace}
\newcommand{\percc}{\ensuremath{\,{\rm cm^{-3}}}\xspace}

\newcommand{\degree}{\ensuremath{^\circ}\xspace}

\newcommand{\arcmper}{\ensuremath{\rlap.{^{\prime}}}}

\newcommand{\jy}{\ensuremath{\,{\rm Jy}}\xspace}
\newcommand{\jyb}{\ensuremath{{\rm \,Jy\,beam^{-1}}}\xspace}

\newcommand{\mjyb}{\ensuremath{{\rm \,mJy\,beam^{-1}}}\xspace}

\newcommand{\te}{\ensuremath{{T_{\rm e}}}\xspace}

\newcommand{\hi}{{\rm H\,{\footnotesize I}}\xspace}
\newcommand{\hii}{{\rm H\,{\footnotesize II}}\xspace}
\newcommand{\cii}{{\rm [C\,{\footnotesize II}]}\xspace}
\newcommand{\nii}{{\rm [N\,{\footnotesize II}]}\xspace}

\newcommand{\hna}{\ensuremath{{\rm H}{\rm n}\alpha\xspace}}
\newcommand{\hnb}{\ensuremath{{\rm H}{\rm n}\beta\xspace}}
\newcommand{\hng}{\ensuremath{{\rm H}{\rm n}\gamma\xspace}}

\newcommand{\form}{\ensuremath{\rm H_2CO}\xspace}
\newcommand{\formr}{\ensuremath{\rm H_2^{13}CO}\xspace}

\newcommand{\methanol}{\ensuremath{\rm CH_3OH}\xspace}

\RequirePackage{rotating}